\documentclass[a4paper,12pt]{JHEP3}

\usepackage{epsfig}

\newcommand{\bc}{\begin{center}}
\newcommand{\ec}{\end{center}}
\newcommand{\be}{\begin{equation}}
\newcommand{\ee}{\end{equation}}
\newcommand{\bq}{\begin{eqnarray}}
\newcommand{\eq}{\end{eqnarray}}

\newcommand{\p}{\varphi}
\newcommand{\Sc}{Schr\"odinger\,}

\title{Solving the Functional \Sc Equation: Yang-Mills String Tension
and Surface Critical Scaling.} 

\author{Paul Mansfield \\Centre for Particle Theory, University of
Durham, Durham DH1 3LE, UK \\ Email: P.R.W.Mansfield@durham.ac.uk}
\abstract{Motivated by a heuristic model of the Yang-Mills vacuum
that accurately describes the string-tension in three dimensions
we develop a systematic method for solving the functional \Sc
equation in a derivative expansion. This is applied to the 
Landau-Ginzburg theory that describes
surface critical scaling in the Ising model. A Renormalisation Group
analysis of the solution yields the value $\eta=1.003$ for the 
anomalous dimension of the correlation function of surface spins
which compares well with the exact result of unity implied by 
Onsager's solution. We give the expansion of the corresponding $\beta$-function
to 17-th order (which receives contributions from up to 17-loops
in conventional perturbation theory).}


\preprint{hep-th/0406237}

\begin{document}

\section{\bf Introduction}

In recent years there have been considerable advances in our
understanding of the quantum field theory of systems with
higher conserved charges. Deep mathematical structures have 
emerged in the study of integrable models in two dimensions
and in supersymmetric theories in higher dimensions. However the 
field theories that are directly applicable to the physical world
do not manifest the high degree of symmetry needed for
mathematical tractability. This limits us to the use
of perturbation theory, which cannot describe strongly interacting
theories, and the largely numerical approach of lattice field theory.
This suggests that it is worthwhile to
develop new methods for the solution of generic non-integrable
quantum field theories. In this paper we will describe
an approach to quantum field theory in which the functional
\Sc equation is constructed in an essentially standard way,
\cite{Hat}-\cite{else},
but is then solved using a derivative expansion for 
the wave-functional from which the continuum limit 
is obtained using a version of the Borel transform.
Like
perturbation theory this yields an approximation
in the form of a series, but unlike perturbation theory 
the method applies at arbitrary values of the coupling.
The approach is inspired by a heuristic model of the Yang-Mills
vacuum but we will develop it for
the toy-model of scalar $\varphi^4$ theory in $1+1$ dimensions.
This is partly because the simplicity of the model allows the
series expansion to be calculated to reasonably high order using
modest computing resources but also because the strong-coupling 
behaviour
of the vacuum wave-functional describes the surface critical
scaling of the Ising model and so we are able to test the method 
against exact results which derive from the Onsager solution and 
conformal field theory.

In a $D$-dimensional gauge theory there is a constant force between static quarks 
leading to confinement when the Wilson loop, $WL[C]$, associated with 
a 
closed curve, $C$, depends on the minimal area spanning $C$ as
$\exp (-\sigma {\rm Area} [C])$, \cite{Wilson}. $\sigma$ is the string tension.
$WL[C]$ is the vacuum expectation value of the trace of the 
path-ordered
exponential of the gauge-field, $WL[C]=\langle 0 |\, {\rm tr}
\,P\,\exp (-\int_{C}A\cdot dx)\,|0\rangle$. This has a functional
integral representation $\int {\cal D}A\,\exp\,(-S_D/g^2)\, 
{\rm tr}
\,P\,\exp\, (-\int_{C}A\cdot dx)$ where $S_{D}$ is the Yang-Mills 
action in $D$ Euclidean space-time dimensions, and $g$ a coupling.
Areal behaviour is observed in Monte
Carlo lattice studies for $D=2,\,3$ and $4$,
but has not been obtained by analytic calculation except in two 
dimensions
where Yang-Mills theory becomes free with the gauge choice $A_{1}=0$.
In 1979 Greensite \cite{Jeff} and Halpern \cite{Halpern} 
proposed a simple model of the confining
Yang-Mills vacuum. For simplicity consider $D=3$. If we choose the 
temporal gauge $A_0=0$ and work in the \Sc representation so that 
$A_{1}$ and $A_{2}$ are diagonal on the quantisation surface $t=0$
then the representative of the
vacuum $\langle A|0\rangle=\Psi[A]$ is a gauge-invariant functional 
of 
$A_{1}(x^{1},x^{2})$ and $A_{2}(x^{1},x^{2})$. The idea is to 
model this three-dimensional vacuum in terms of the two-dimensional
action as $\Psi\approx \exp\,(-kS_{2}/g^{4} )$, with $k$ a 
dimensionless constant. This guarantees 
gauge-invariance and if we choose a planar $C$ so that it
can be rotated into
the quantisation surface then the usual expression of vacuum 
expectation values within the \Sc representation yields 
\bq
WL[C]=&&
\int {\cal D}A_{1}\,{\cal D}A_{2}\,\Psi^{*}[A]\,{\rm tr}
\,P\,\exp\, (-\int_{C}A\cdot dx)\,\Psi[A]\nonumber\\
=&&\int {\cal D}A_{1}\,{\cal D}A_{2}\,\exp(-2kS_2/g^{4})\, {\rm tr}
\,P\,\exp\, (-\int_{C}A\cdot dx)
\eq
which reduces to the Wilson-loop for a two-dimensional
gauge theory for which we can compute the string tension
in terms of $k$. This form of $\Psi[A]$ was compared
directly to Monte Carlo
simulations of the vacuum functional in lattice gauge theory in
\cite{Iwa}. It might now be timely to repeat this comparison
given the improvements in lattice gauge theory technology.
This argument was generalised to quantum gravity in 
\cite{Horiguchi2}-\cite{Maeda}.

$\Psi[A]$ should satisfy the \Sc equation, and this has been used to
estimate $k$, \cite{Paul}. The Hamiltonian is $-\int d^2x\,{\rm tr}\,(\dot A^{2}
+B^{2})$ where the `magnetic field', $B$, is the $1,2$-component of 
the field strength from which the two-dimensional action is 
constructed
as $S_{2}=-\int d^2x\,{\rm tr}\,(B^{2})$. In the \Sc representation
$\dot A=-ig^{2}\delta /\delta A$, so that the \Sc equation is
\be
H\,\Psi=\int d^2x\,{\rm tr}\,\left(g^2{\delta^2\over\delta A^{2}}
-{1\over g^{2}}B^{2}\right)\,\Psi=E\,\Psi\,.
\ee
Differentiating $S_{2}$ functionally with respect to $A$ gives
the covariant derivative of $B$, $D_{i}B$. Differentiating again
gives a second order partial differential operator, $\Omega$,
so that for the model vacuum the \Sc equation becomes
\be
H\,\Psi=\left( {k\over g^{2}}{\rm Tr}\,\Omega+
{1\over g^6}{k^{2}}\int d^2x\,{\rm tr}\,((DB)^{2})+
{1\over g^{2}}S_2\right)\Psi=E\,\Psi\,.  \label{apsc}
\ee
When the functional trace of $\Omega$ is regulated
using a heat-kernel based regulator to preserve gauge invariance
the result is a divergent constant plus a multiple
of $S_{2}$, so that $k$, and hence $\sigma$, are determined if we 
ignore $(DB)^{2}$ as a higher order 
correction. Specifically,
if zeta-function regularisation \cite{Paul} is used we obtain
\be
\sigma={23g^4(N^2-1)\over 192\pi}\equiv\sigma_\zeta \label{st}
\ee
for the gauge-group $SU(N)$, (and where we have adapted the result of \cite{Paul}
to the convention ${\rm tr} \,(T_{a}T_{b})=-\delta_{ab}/2$ for the 
generators of the fundamental representation.) This formula
shows remarkable agreement with recent lattice studies.
Figure (\ref{strintt}) plots $\sqrt\sigma/(g^{2}N)$ against $N$ as measured in 
Monte Carlo
simulations \cite{Teper}. $\sqrt\sigma_\zeta/(g^{2}N)$
is shown
on the same figure, and the values agree to within a half
to one percent which is significant given that the formula (\ref{st})
has no adjustable parameters. (There is a similar formula, 
$\sigma_{KKN}=g^{4}(N^2-1)/(8\pi)$, \cite{Nair},based on a different parametrisation
of the canonical variables which is also displayed in the Figure).
\FIGURE{\centerline{\epsfysize=11cm\epsfbox{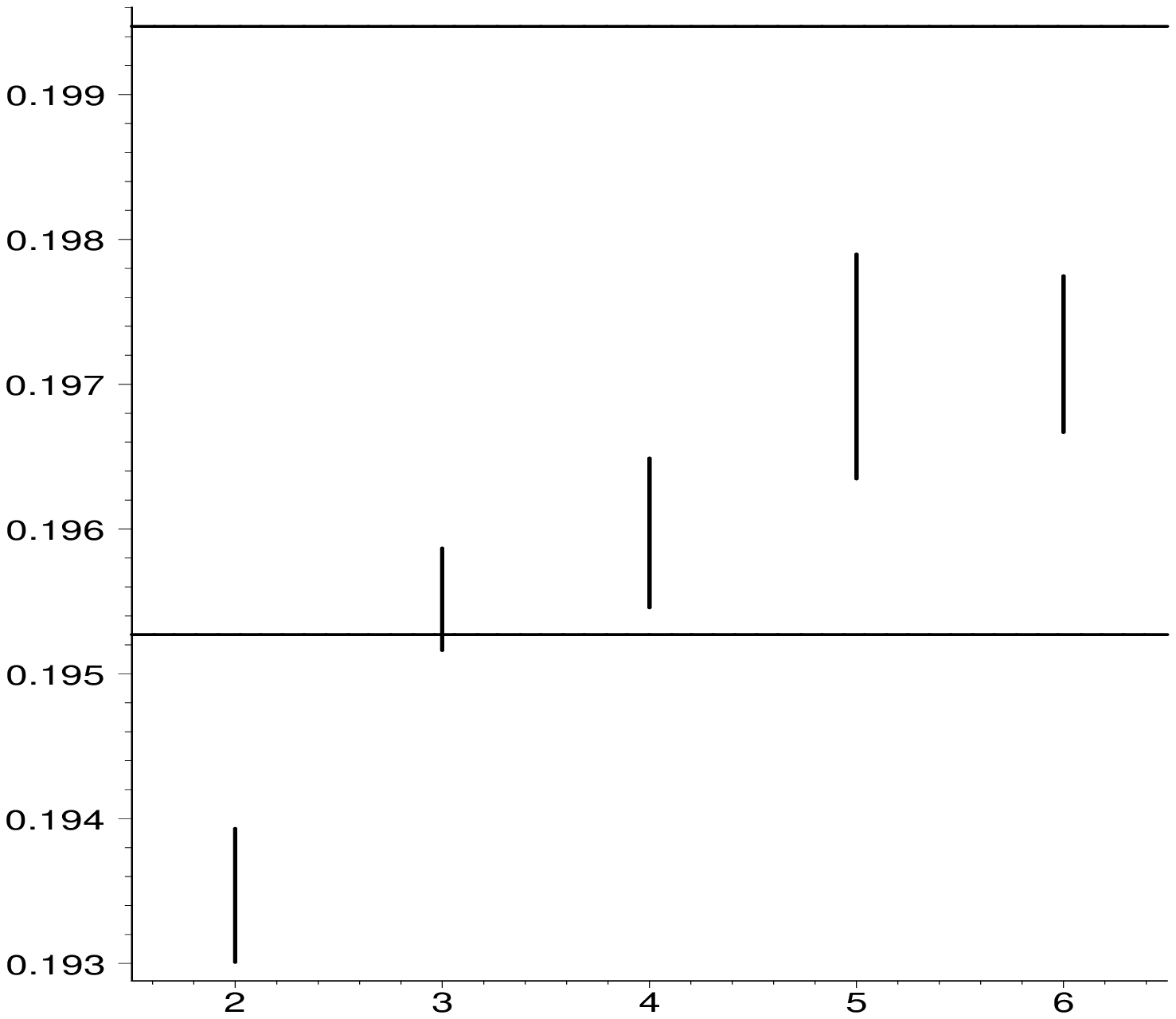}\caption[series]{{\label{strintt} $\sqrt\sigma/(g^{2}\sqrt{N^2-1})$ against $N$
from lattice QCD, \cite{Teper}.
The horizontal lines correspond to the predictions from  
$\sigma_\zeta$ (lower) and
$\sigma_{KKN}$.
}}}}

This model begs many questions:
Can the vacuum functional really be 
the 
exponential of the integral of a local quantity, i.e. depending on a 
single
point in space, since this is quite unlike the non-local functionals 
seen in 
perturbation theory. For example the free Maxwell field in 
four-dimensions has 
$$\Psi=\exp (-\int d^{3}x\,d^{3}y\,B(x)\,B(y)/(e^{2
}|x-y|^{2}))\,.$$
Is the approach regularisation dependent, and
is it consistent with the usual ultra-violet behaviour and
renormalisation of gauge theories? 
Is the model Lorentz invariant?
Can corrections to the model be computed  and 
shown to converge to the lowest energy solution to the \Sc equation?
Applying the argument to four-dimensional gauge theory would
give a vacuum functional that is the exponential of the three-dimensional
action with a cut-off dependent coefficient that
appears to lead to a divergent string tension.
Despite these issues 
the agreement between lattice simulations and the formula for the 
string tension suggests that it is worthwhile
to investigate the approach further and make it more systematic.

In calculating $k$ the contribution of $(DB)^{2}/g^{6}$ was ignored in 
(\ref{apsc}).
This may be corrected for by including a similar term in
the vacuum functional. This leads to an adjustment to the coefficient
of the original term proportional to ${\rm tr} (B^2)$ because $\int 
d^{2}x\,\delta^{2}/\delta A^{2}$
applied to $\int 
d^2x\,{\rm tr}((DB)^{2})$ 
gives a piece proportional to $\epsilon^{-2}\int d^{2}x {\rm tr}\, 
B^{2}$,
where $\epsilon$ is  a short-distance cut-off
as well as $\int 
d^2x\,{\rm tr}((DB)^{2})$ 
so that now, if we set $\Psi=\exp W$ then the approximation for $W$ is
$$
W\approx \int d^2x\,{\rm tr} \left(\left({k\over g^{4}}+{k_{3}\over 
\epsilon^2
g^{8}}\right)B^{2}+{k_{2}\over g^{8}}(DB)^{2}\right)\,,\label{imp}
$$
which satisfies the \Sc equation for appropriate values of $k_2$ and 
$k_3$ if now we ignore terms of mass dimension eight.
These higher dimension expressions can then be taken into account by 
adding further terms so that the process generates 
an expansion for $W$ consisting of local expressions of increasing 
dimension 
with coefficients that are themselves series in decreasing powers of 
the cut-off $\epsilon$. Superficially such a derivative expansion 
would appear to 
be applicable only to slowly varying fields, making it difficult to
address the ultra-violet behaviour that has to be understood
in order to renormalise the theory. Furthermore the cut-off
dependence of the coefficients is unexpected since a theorem
due to Symanzik \cite{Sym} implies that $W$ should be finite as 
$\epsilon\rightarrow 0$ (at least to all orders in
perturbation theory).

To address the problem of whether a systematic expansion of the 
vacuum 
functional can be obtained as a derivative expansion with the correct
ultra-violet behaviour we turn to a simpler model than gauge theory,
namely scalar $\p^{4}$  theory in two space-time dimensions. The 
simplicity of the model will allow us to pursue the expansion to high 
order and, as stated earlier, the strong-coupling ultra-violet
behaviour is related to the surface critical scaling of the Ising 
model,
providing exact results to test our approach against.

The Hamiltonian for scalar $\p^{4}$ in two space-time dimensions
is
\be
H=\int dx \left( {1\over 2}\left (\dot\p_0^{2}+\p_0^{\prime 2}
+m_{0}^{2}\,\p_0^{2}\right)+{g_{0}\over 4!}\,\p_{0}^{4}\right)\,,
\ee
where $\p_{0}$, $m_{0}$ and $g_{0}$ are the bare
field, mass and coupling. Quantisation consists simply of
replacing $\dot\p_0$ by $-i\delta/\delta\p_{0}$.
We can now look for eigenfunctionals of $H$. 
We will assume that, as in quantum mechanics, 
the vacuum functional has no 
nodes, so that it can be written as $\Psi=\exp W$,
giving a \Sc equation
\be
\int dx \left(-{\delta^{2}W\over\delta\p_{0}^{2}}
-\left({\delta W\over\delta \p_{0}}\right)^{2}
+\p_0^{\prime 2}+m_0^{2}\,\p_{0}^{2}+{g_{0}\over 12}\,\p_{0}^{4}
\right)=E_{v}\,\label{barese}\ee
where $E_{v}$ is the energy of the vacuum.
The first term needs to be regularised, and we adopt a
simple momentum cut-off, replacing 
${\delta^{2}/\delta\p_{0}^{2}}$
by
\be
{\delta^{2}\over\delta\p_{0}^{2}}\Big |_{s}
\equiv
\int dx\,dy\,\int_{|k|<1/s}{dk\over 2\pi} e^{ik(x-y)}
{\delta^{2}\over\delta\p_{0}(x)\,\delta\p_{0}(y)}
\ee
with $s$ a small distance. 

The problem of renormalisation is to
tune the bare quantities in terms of $s$ to obtain
finite physics. Renormalisation conditions have also to be chosen
to replace the (potentially divergent) bare quantities by finite
parameters, in this case a mass scale $\mu$ and dimensionless
coupling $g$.
We are aided in this by Symanzik's theorem \cite{Sym}
which implies that provided we renormalise $\p_{0}$
multiplicatively $\p_{0}=\sqrt Z \,\p$ (and tune $m_0$ and $g_0$)
then $W$ is finite apart from
a local term proportional to $\int dx\,\p^{2}$.

Conventional semi-classical perturbation theory can be
developed \cite{Hat} by re-instating $\hbar$ in the quantisation 
$\dot\p_0=-i\hbar\delta/\delta\p_{0}$ and expanding
$W=W_{1}/\hbar+W_{2}+\hbar W_{3}+..$ so that to leading order
(\ref{barese}) becomes the Hamilton-Jacobi equation
with a solution most conveniently expressed in terms
of the Fourier transform $\tilde \p$ as
\be
W_{1}=-\int {dp\over 4\pi}\tilde\p (-p)\omega (p)
\tilde\p (p)
-{g_0\over 4!}\int{dp_{1}\over 2 \pi}..{dp_{4}\over 2 \pi}
2\pi\delta(p_1+..+p_{4})\,{\tilde\p(p_{1})..\tilde\p(p_{4})\over 
\omega(p_1)+..+ \omega(p_4)}
+..
\ee
where $\omega=\sqrt{p^2+m_{0}^{2}}$, and bare and renormalised 
quantities coincide at this (tree) level. Clearly for
slowly varying $\p$ we can take $|p|<<m_{0}$
and expand $\omega=m_0+p^2/(2m_{0})+..$ to obtain a local
expansion.  The one-loop, i.e. order $\hbar$, correction can be readily
obtained, \cite{Hat}, \cite{Marcos}, but this is not our purpose here. Instead we
will develop the approach of \cite{Paul1} and consider solving for the full $W$ starting from an expansion  
in a basis of local functionals of the renormalised field
with coefficients $\{b\}$:
\be
W=\int dx\left(b_2\p^2+b_4\p^4+b_6\p^6+..
+b_{2,2}\p^{\prime2}+b_{4,2}\p^{\prime2}\p^2+..+b_{2,4}\p^{\prime\prime 
2}+..
\right)\,
\label{localexp}\ee
Now the regulated second functional derivative acts on this
local expansion to produce another local expansion, for example
\be
\int dx\,{\delta^{2}\over\delta\p^{2}}\Big |_{s}
\,\int dy\, \p^2\,\p^{\prime 2}=\int dx\,\left({2\over\pi s}\,\p^{\prime 
2}+{2\over 3\pi s^3}\,\p^2 \right)\,.
\ee
The other terms in the \Sc equation are also local functionals of
$\p$ so we can reduce this functional differential equation 
to an infinite set of algebraic equations for the infinite number of
constants
$\{b\}$ by equating coefficients of the basis functionals to zero.
To solve these requires some form of truncation, for example 
by simply ignoring coefficients of local terms containing more than a
prescribed number of fields or derivatives.
In \cite{Paul1} a different truncation was proposed that
would use an estimate of such `high order' coefficients.
Once the problem has been made finite 
the equations can be solved, for example numerically,
although the stability and convergence of the solutions
may be sensitive to the truncation.
We will use a different approach to that of \cite{Paul1}.
Rather than truncate the expansion (\ref{localexp})
we look for solutions to the full set of 
(untruncated) equations in the form of 
expansions of the coefficients $\{b\}$ as power series in $1/s$.
We will see that the enlarged set of equations obtained by equating to 
zero the various powers of $s$ form closed systems that
can be solved rapidly and simply in terms of the renormalised
quantities $\mu$ and $g$. 
The quantities
$Z,g_0$ and $m_0$ will similarly be obtained as power series in
$1/s$. We will compute these series up to some 
fixed order, $N$, determined by computing resources.
This is the only truncation we will use.

The continuum limit is obtained by sending $s$ to zero, (and $N$ to infinity)
so these series solutions would appear to have little physical 
relevance, as they are only likely to converge for large $s$.
However the small $s$ behaviour can be extracted using a version of the Borel transform,
which we describe in Section 3. 

\FIGURE{\centerline{\epsfysize=9cm\epsfbox{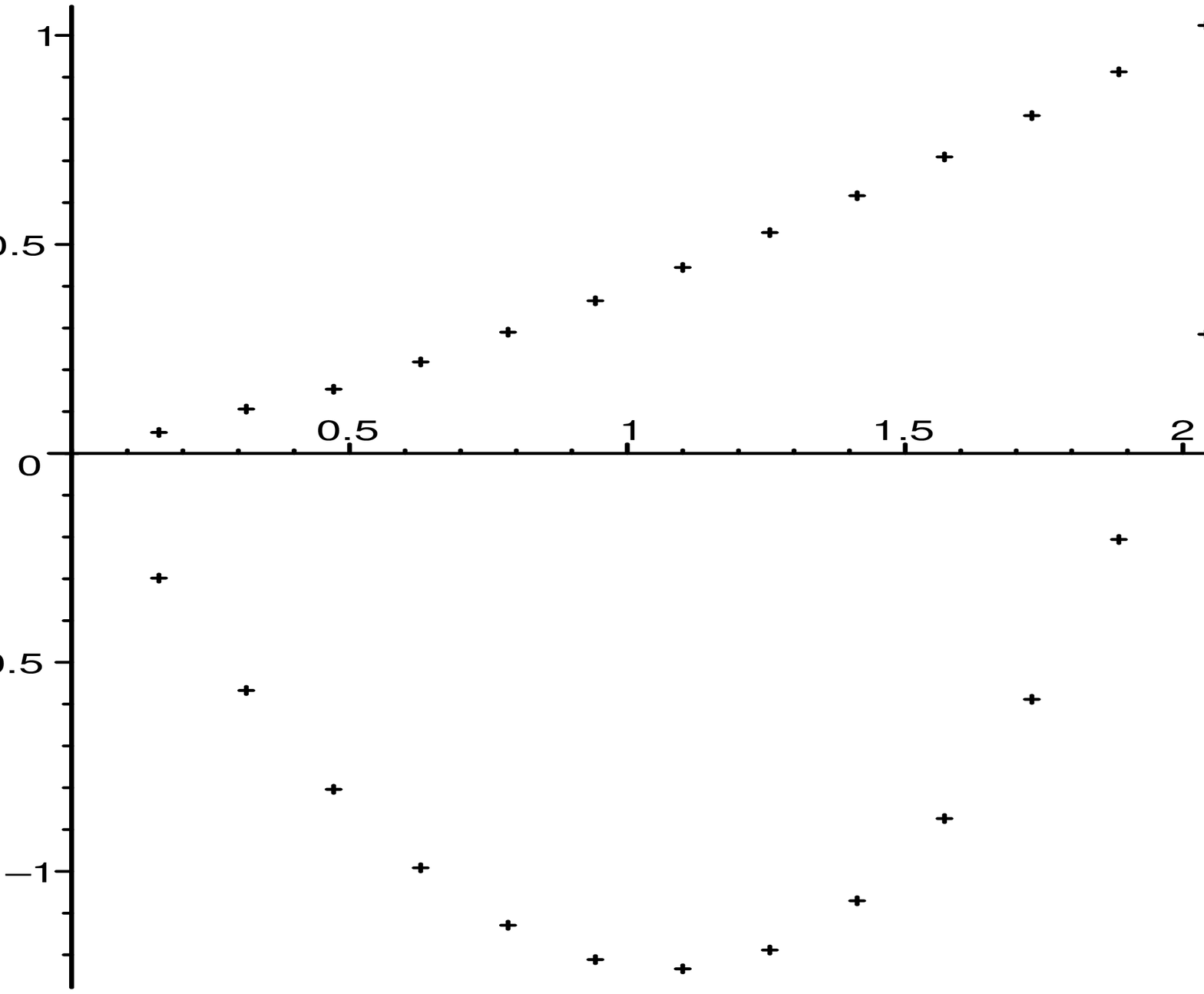}\caption[series]{\label{resumex5} $\gamma (g)$ (upper curve) and $\beta(g)$
from the \Sc equation.}}}

We will test our solution of the \Sc equation against exact
results obtained for the surface critical scaling of the Ising model 
from the Onsager solution and conformal field theory \cite{Cardy}.
At a critical value of the renormalised coupling, $g$, the 
$\mu\rightarrow 0$ limit of $\p^4$ theory describes the Ising model
at criticality. More specifically the wave-functional describes
the Ising model with a boundary (corresponding to the quantisation 
surface) 
at the `Ordinary Transition'. That part of the
wave-functional that is quadratic in $\p$ is related to the
correlator of two spins lying in the boundary
which is known to depend on their separation $|x|$
as $1/|x|$. This quadratic piece can be written as $\int dx\,dy\, 
\p(x)
\p(y)/|x-y|^{2-\eta}$ where $\eta$ is the anomalous dimension,
so the correspondence with the Ising model requires that $\eta=1$.
Now a standard Renormalisation Group argument presented in Section 4
gives $\eta$ as the value of $d \log Z /d\log \mu\equiv \gamma (g)$ 
computed at the zero of the beta-function $d g/d\log \mu\equiv \beta(g)$. 
In Figure \ref{resumex5}
we display these two functions computed from our solution when we 
truncate the series expansions to order $1/s^{17}$ terms.
The graph
indicates good agreement with the exact result, and as we will
see the agreement is considerably improved upon extrapolation in the 
number of terms, $N$.

\section{Solving the \Sc Equation for $\p^{4}$.}

The \Sc equation for the vacuum, $H \exp W=E_v\exp W$, can be written 
in terms of the renormalised field as
\be
\int dx\left({\delta^{2}W\over\delta\p^{2}}\Big |_{s}+
\left({\delta W\over\delta\p}\right)^{2}-Z^{2}\p^{\prime 
2}-m_0^{2}Z^{2}\p^{2}
-{g_{0}\over 12}Z^{3}\p^{4}+2Z{\cal E}_{v}\right)=0
\ee
where ${\cal E}_{v}$ is the energy density of the vacuum.
The local expansion (\ref{localexp}) takes the form
\bq
W=\int dx
\Big(&&b_2\,\p^2+b_4\,\p^4+b_6\,\p^6+..
+b_{2,2}\,\p^{\prime2}+b_{2,4}\,\p^{\prime\prime 2}+..\nonumber\\
&&+b_{4,2}\,\p^{\prime 
2}\,\p^2+b_{4,4,1}\,\p^{\prime\prime 
2}\,\p^2 +b_{4,4,2}\,\p^{\prime 4}
+..\nonumber\\
&&+b_{6,2}\,\p^{\prime2}\,\p^4+..+b_8\,\p^8+b_{8,2}\,\p^{\prime2}\,\p^6+..
\Big)\,\label{locexp2}
\eq
where the dots denote an infinite number of 
terms containing increasing powers of the field and its derivatives.
The two (or three) labels on the coefficients denote the number of fields, 
the number of derivatives (and a further label to distinguish between
different terms for which the first two labels are the same).
We will not truncate the expansion at this stage.
Substituting (\ref{localexp}) into the \Sc equation gives
\bq
\int dx
\Big(&&\Big[\left( {2b_{2}\over\pi s}+ {2b_{2,2}\over3\pi 
s^{3}}+..\right)
+\p^{2}\left({12 b_{4}\over\pi s}+{2 b_{4,2}\over3\pi s^{3}}+..\right)\nonumber\\
&& 
+\p^{\prime 2}\Big({2 b_{4,2}\over\pi s}+{4 b_{4,4,1}\over \pi s^{3}}
+{8 b_{4,4,2}\over 3\pi s^{3}}+..\Big)
+\p^{4}\left({30 b_{6}\over\pi s}+{2 b_{6,2}\over3\pi 
s^{3}}+..\right)\nonumber\\
&& 
+\p^{6}\left({56 b_{8}\over\pi s}+{2 b_{8,2}\over3\pi s^{3}}+..\right)
+\p^{2}\p^{\prime 2}\left({12 b_{6,2}\over\pi 
s}+..\right)+..\Big]\nonumber\\
&&+\Big[\p^{2}4 b_{2}^{2}
+\p^{\prime 2} 8b_{2}b_{2,2}+\p^{4}16 b_{2}b_{4}+\p^{6}\left(24 
b_{2}b_{6}
+16b_{4}^{2}\right)+\nonumber\\
&&
\p^{2}\p^{\prime 2}\left(16 b_{2}b_{4,2}+48 
b_{2,2}b_{4}\right)+..\Big]
\nonumber\\
&&
-Z^{2}\p^{\prime 2}-m_0^{2}Z^{2}\p^{2}
-{g_{0}\over 12}Z^{3}\p^{4}+2Z{\cal E}_{v}\Big)=0\label{locSW}
\eq
We can also look for an eigenstate of the Hamiltonian describing a 
single particle at rest in the form $\Psi=U\,\exp W$ 
The \Sc equation for this state, $H \Psi=(\mu+E_v)\Psi$,
can be written as
\be
\int dx\left({\delta^{2}U\over\delta\p^{2}}\Big |_{s}+
2{\delta W\over\delta\p}{\delta U\over\delta\p}+2\mu Z U\right)=0
\ee
Take  
$U$ as another local expansion:
\be
U=\int 
dx\Big(\p+c_{3}\p^{3}+c_{5}\p^{5}+..+c_{3,2}\p\p^{\prime 
2}+c_{3,4}\p\p^{\prime\prime 
2}+..+c_{5,2}\p^{3}\p^{\prime 
2}+..\Big)\,,\label{locexpU}
\ee
where again the dots denote an infinite number of local terms
all of which we keep in the expansion for the time being.
Substituting into the \Sc equation gives
\bq
\int dx
\Big(&&\Big[\p\left( {6c_{3}\over\pi s}+ {2c_{3,2}\over3\pi 
s^{3}}+{2c_{3,4}\over 5\pi s^{5}}+..\right)
+\p^{3}\left({20 c_{5}\over\pi s}+{2 c_{5,2}\over3\pi s^{3}}+..\right)\nonumber\\
&&
+\p\p^{\prime 2}\left({6 c_{5,2}\over\pi s}+..\right)+..\Big]
\nonumber\\
&&+2\Big[\p 2 b_{2}
+\p^{3} \left(b_{4}+6c_{3}b_{2}\right)+
\p\p^{\prime 2}\left(12 c_{3}b_{2,2}+6 
c_{3,2}b_{2}+2b_{4,2}\right)+..\Big]
\nonumber\\
&&
+2\mu  
Z\left(\p+c_{3}\p^{3}+c_{5}\p^{5}+..+c_{3,2}\p\p^{\prime 
2}+c_{3,4}\p\p^{\prime\prime 
2}+..\right)  \Big)=0\label{locSU}
\eq
These equations have to be solved in conjunction with
renormalisation conditions that identify parameters that remain
finite as the cutoff is removed. We will take $\mu$ to be our finite
mass-scale. Symanzik's theorem \cite{Sym} implies that the coefficients $b$ 
in the local expansion of $W$ are finite, with the possible exception
of $b_{2}$, so we choose renormalisation conditions
\be
b_{4}=-{g\mu\over 8},\quad b_{2,2}=-{1\over 4\mu}\,,\label{rencon}
\ee
with dimensionless $g$ playing the r\^ole of renormalised coupling.
Our aim is to solve (\ref{locSW}) and (\ref{locSU})
for the coefficients
$b$ and $c$, and also for $g_{0}$, $m_{0}$ and $Z$ in terms of the
finite parameters $\mu$ and $g$.

Equating the coefficients of the basis functionals to zero in
(\ref{locSW}) gives an infinite set of algebraic equations.
This may be divided into two classes. In the first class
are the equations obtained from the coefficients
of $1$, $\p^{2}$, $\p^{\prime 2}$,  and $\p^{4}$.
\be
{2b_{2}\over\pi s}+ {2b_{2,2}\over3\pi 
s^{3}}+..+2Z{\cal E}_{v}=0,
\quad {12 b_{4}\over\pi s}+{2 b_{4,2}\over3\pi s^{3}}+..+4 
b_{2}^{2}-m_0^2Z^2=0
\ee
\be{2 b_{4,2}\over\pi s}+{4 b_{4,4,1}\over \pi s^{3}}
+{8 b_{4,4,2}\over 3\pi s^{3}}+..+8b_2b_{2,2}-Z^2=0,\quad 
{30 b_{6}\over\pi s}+{2 b_{6,2}\over3\pi 
s^{3}}+..+16b_2b_4-{g_{0}\over 12}Z^{3}=0\label{bareq}
\ee
These equations involve the bare quantities $g_{0}$, $m_{0}$, $Z$
and ${\cal E}_{v}$ as well as the coefficients $b$.
The second class consists of all the other equations,
and these involve only the coefficients $b$.
\be
{56 b_{8}\over\pi s}+{2 b_{8,2}\over3\pi s^{3}}+..
+24 b_{2}b_{6}+16b_{4}^{2}=0,\quad
{12 b_{6,2}\over\pi 
s}+..
+16 b_{2}b_{4,2}+48 
b_{2,2}b_{4}=0.\label{beq}
\ee
\be \ldots\quad \quad \quad \quad \quad \quad \quad \quad \quad \quad \quad \quad \ldots\ee

We begin by solving this second class for the $b$
coefficients other than $b_{2}$, $b_{4}$ and $b_{2,2}$
in terms of $g$, $\mu$ and $b_{2}$ using (\ref{rencon}). Expanding
each coefficient as a power series, 
$b_{n}=\sum_{r=0..\infty}b_{n}^{r}/s^{r}$,
and then equating the coefficients of powers of 
$1/s$ to zero in each equation gives further equations
that can be easily solved in MAPLE to high order.
An important feature is that they form
a closed system which is easily solved order by order in $1/s$.
Consider the equation corresponding to the coefficient
of a given local term involving $F$ powers of the field
and a total of $D$ derivatives.
The equation obtained by equating to zero 
the coefficient of $s^{-p}$ takes the form: 
\bq
&&\sum_{D'=D}^{D+p-1}  A^{p,D',r'}_{F,D,r}\,b_{F+2,D',r'}^{p-1+D-D'}
+b_2\,b_{F,D,r}^p
\nonumber\\
&&+\sum_{D''=0,\,F'=4,\,p'=0\,,r'\,,r''}^{D''=D,\,F'=F-4,\,p'=p}
C^{p,D'',F',p',r',r''}_{F,D,r}\,b_{F',D'',r'}^{p'}
\,b_{F+2-F',D-D'',r''}^{p-p'}=0
\eq
where $A$ and $C$ are known numerical coefficients.
This can be solved for $b_{F,D,r}^p$ in terms of the coefficients
$\{b_{F',D',r'}^{p'}\}$ where $F'+2D'+2p'\le F+2D+2p\equiv Q$
and either $F'=F$, $p'<p$ or $F'<F$, $p'\le p$,
$D'\le D$. 
\bq
b_{F,D,r}^p=
&&-{1\over b_2}\Big(\sum_{D'=D}^{D+p-1}  A^{p,D',r'}_{F,D,r}\,b_{F+2,D',r'}^{p-1+D-D'}
\nonumber\\
&&+\sum_{D''=0,\,F'=4,\,p'=0\,,r'\,,r''}^{D''=D,\,F'=F-4,\,p'=p}
C^{p,D'',F',p',r',r''}_{F,D,r}\,b_{F',D'',r'}^{p'}
\,b_{F+2-F',D-D'',r''}^{p-p'}\Big)
\eq
To obtain $b_{F,D,r}^p$ explicitly we
can build up the solutions 
starting with $p=0$.
Given that the renormalisation
conditions fix $b_4^0$ and $b_{2,2}^0$ we begin by calculating
$b_6^0$, $b_{2,4}^0$ and $b_{4,2}^0$ and then use these results
to calculate $b_8^0$, $b_{4,4}^0$, $b_{6,2}^0$ and $b_{2,6}^0$.
From these we compute more coefficients
until all the $\{b_{F',D'}^{0}\}$ are known
with $F'+2D'\le Q$. Then we move on to $\{b_{F',D'}^{1}\}$
by first using the results obtained to compute
$b_6^1$, $b_{2,4}^1$ and $b_{4,2}^1$. From these we obtain
$b_8^1$, $b_{4,4}^1$, $b_{6,2}^1$ and $b_{2,6}^1$
and continue the process until all the $\{b_{F',D'}^{1}\}$ are known
with $F'+2D'+2\le Q$, and from these we can compute
$\{b_{F',D'}^{2}\}$ and so on. Although this appears complicated
in practice it is straightforward.
For example (\ref{beq}) gives to lowest order in $1/s$
\be
24 b_{2}b_{6}^{0}+{(g\mu)^{2}\over 4}=0\Rightarrow b_{6}^{0}
=-{(g\mu)^{2}\over 96b_2},\quad
16 b_{2}b_{4,2}^{0}+ 
{3g\over 2}=0\Rightarrow b_{4,2}^{0}=-{3g\over 32b_2}.
\ee

Once the $\{b_{F,D}^{p}\}$ are known
substituting the solutions into the first class of equations
yields series for the bare quantities.
Thus the first of (\ref{bareq})
gives
\be
Z^{2}=-{2b_2\over \mu}-{3g\over 16 b_{2 }\pi s}+..
\ee
Having obtained the $b$-coefficients and $Z$, $m_0$ and $g_0$
we solve (\ref{locSU})
to obtain the $c$-coefficients and the energy eigenvalue
in terms of $\mu$, $g$ and $b_{2}$ again as power series in $1/s$. Finally,
given that the energy eigenvalue equals $\mu$
we can invert the series to obtain $b_{2}$
in terms of $g$ and $\mu$, and substitute this into all
the other series expansions to express all quantities
as power series in $1/s$ depending on $g$ and $\mu$.
In the Appendix we give the expansions \footnote{We note in passing that 
such series could be obtained
from standard Feynman diagram 
expanded in powers of $1/s$, for large $s$.
Given that these expressions contain powers of the coupling to order
$g^{17}$ and $g^{18}$ this would require working to 17 loops in conventional 
perturbation theory.} for $g_{0}$,
and $Z$ up to order
$1/s^{17}$ which requires taking terms with $F+2D$ up to 40 in the local
expansions (\ref{locexp2}) and (\ref{locexpU}), of which there are about 625
and 540 respectively.

These expansions can only be expected to converge for 
large values of $s$, if at all, but the continuum limit
requires taking $s$ to zero. In the next section we will show how
a version of the Borel transform can be used to extract the small $s$ 
behaviour
from these series, and then in Section 4 we will
apply this to the bare quantities to extract an anomalous scaling 
dimension which we can test against the exact result
for the Ising model.

\vfill
\eject

\section{Small $s$ from large $s$ via Borel}

The functional integral representation of wave functionals
described in section 4 can be expanded in terms of
Feynman diagrams. For massive field theories
Feynman diagrams are analytic in
the cut-off $s$ for large enough $s$, although for small
$s$ they have singularities in the left-half-plane.
After renormalisation they have finite limits both as $s\rightarrow 0$
and as $s\rightarrow \infty$ (where they tend to their tree-level 
values). Assuming that these properties hold 
beyond perturbation theory
we will exploit this analyticity to 
extract the small-$s$ behaviour from our series expansions.

As an example, consider a function $f(s)$ that has the assumed 
analyticity properties of wave-functionals, namely it is analytic for 
large $s$, and in the right-half-plane, with finite limits as
$s$ approaches the origin and infinity. For large $s$ it will have a
Laurent expansion containing only non-positive powers of
$s$, but suppose that we only know the first few terms,
$A_N(s)=\sum_{n=0..N}a_{n}/s^{n}$, from which we wish to estimate 
the small $s$ behaviour.
We can obtain this from the integral
\be
I^N_\lambda={1\over 2\pi i}\int_{C} ds{e^{\lambda(s-s_{0})}\over
s-s_{0}}A_N(s)
\ee
where $C$ is a large circular contour centred on the origin
running from just below the negative real axis to just above.
This integral is readily computable in terms of the coefficients
$a_{n}$. Also, since $A_N(s)\approx f(s)$ on $C$:
\be
I^N_{\lambda}\approx{1\over 2\pi i}\int_{C} 
ds{e^{\lambda(s-s_{0})}\over
s-s_{0}}f(s)\equiv I_\lambda\,.
\ee
If for $I_\lambda$ the contour $C$ is now collapsed to surround the singularities of
the integrand, then the pole at $s_0$ contributes $f(s_0)$.
Poles and cuts lying to the left of $s_0$ will be suppressed
when ${\rm Re} (s_0)>0$, by factors exponential in $\lambda$.
For example, a pole at $s=s_p$ contributes a piece proportional to
$\exp (\lambda(s_p-s_0))$. Similarly the size of the contribution
from a contour, $C_f$, of finite length extending as far to the right 
as 
$s_c$ is bounded by
\be
|{1\over 2\pi i}\int_{C_f} ds{e^{\lambda(s-s_{0})}\over
s-s_{0}}f(s)|<e^{{\rm Re}(\lambda(s_c-s_{0}))}
{1\over 2\pi }\int_{C_f} {|ds|\,|f(s)|\over
|s-s_{0}|}
\ee
This exponential damping means that for large $\lambda$
\be
I^N_{\lambda}\approx f(s_0)\,.
\ee
We cannot take $\lambda$ arbitrarily large, but only 
as large as allows $I^N_\lambda$ to be a good approximation
to $I_\lambda$, which is the case when $I^{N-1}_\lambda$
differs by no more than say a few per cent from $I^N_\lambda$.
If this condition fixes $\lambda$ to be $\lambda*$ then
we arrive at the approximation $f(s_0)\approx I_{\lambda*}^{N}$

We will illustrate this with the truncated series
$A_N=\sum_{n=0}^N (-1/s)^n/(n+1)$. As $N\rightarrow \infty$
this series converges for $|s|>1$ to $f=s\log (s/(s+1))$,
which is analytic except for a cut from $s=-1$ to the origin,
so
\be
I_\lambda=s_0 \log \left({s_{0}\over s_0+1}\right)+\int_0^1 dy {y\over y+s_0}
e^{-\lambda(y+s_0)}
\ee
from which it is clear that $\lim_{\lambda\rightarrow\infty}
I_\lambda= f(s_0)$. If we use the series truncated at
say nineteen terms then $A_{19}(s)$ itself approximates $f(s)$
to better than one percent for $s>1.07$. At $s=1$ the error
grows to three and a half per cent and for $|s|<1$ the error
increases rapidly as $s$ gets smaller. Figure \ref{resumex1} plots $I_\lambda^{18}$
and $I_\lambda^{19}$ for $s_0=0.9$. 
\FIGURE{\centerline{\epsfysize=9cm\epsfbox{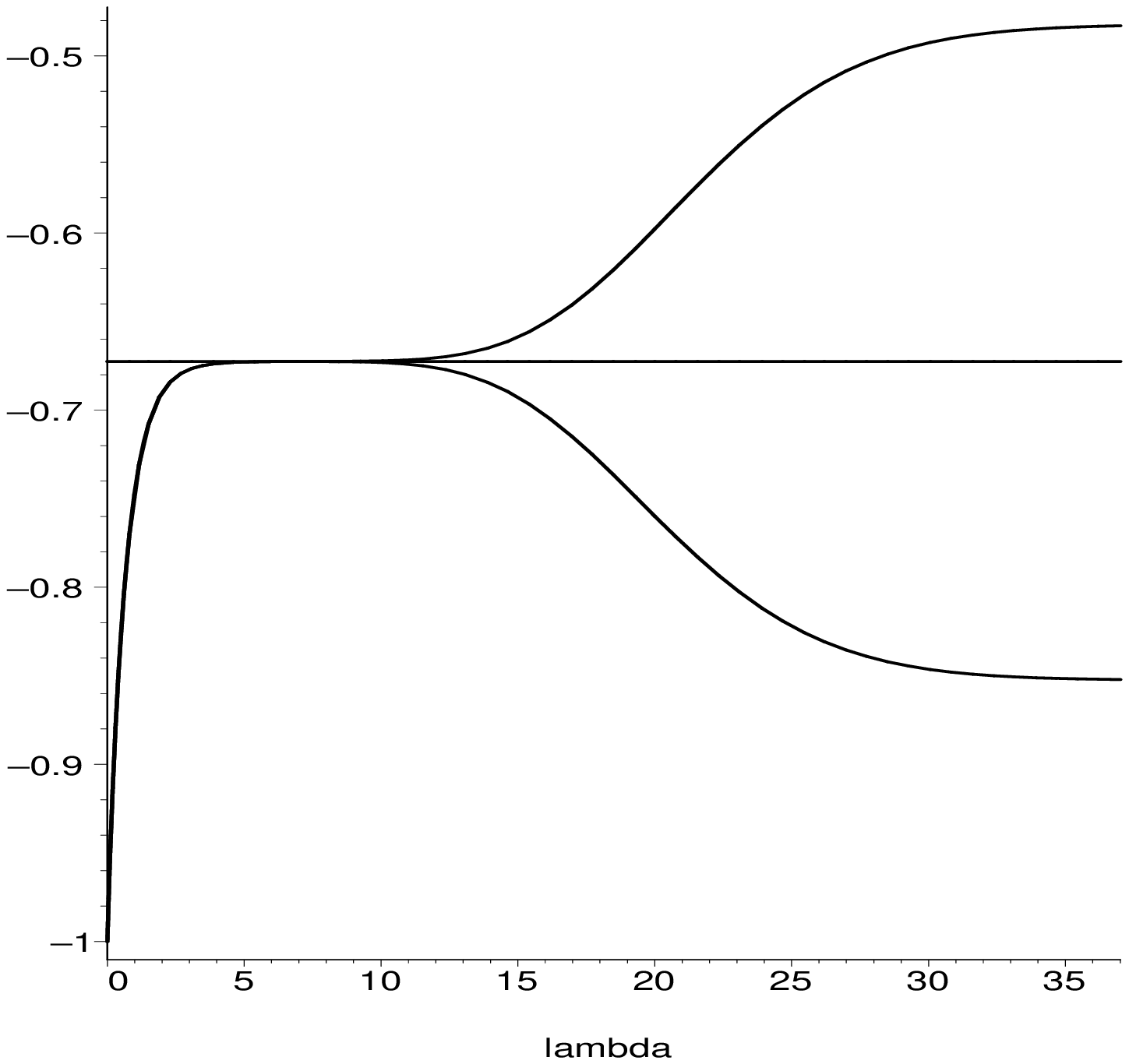}\caption[series]{\label{resumex1} $I_\lambda^{18}$ (lower curve) and $I_\lambda^{19}$}}}
The two curves agree
up to about $\lambda=9$, and for this value provide a very good 
approximation 
to $f(s_0)=-0.6725$, ($I_\lambda^{19}=-0.6724$ and 
$I_\lambda^{18}=-0.6727$). For larger values the two curves diverge, 
and 
approach  $A_{19}(0.9)=-0.4823$ and $A_{18}(0.9)=-0.8524$
which are poor approximations to $f(0.9)$.
For much smaller values of $s_0$ we still obtain reasonable 
results, for example in the case of $s_0=0.05$
the two curves diverge at $\lambda\approx 7$ yielding an 
approximation of $I_\lambda^{19}=-0.2123$ to $f(0.05)=-0.1522$,
which is fifty per cent out, but it is amusing to compare this 
to the value of $A_{19}(0.05)=0.25\times 10^{24}$.

We will be principally interested in the value of $f$ at the origin. 
We can take $s_0\rightarrow 0$ in the general case,
but the new feature that arises is that any cut in the left 
half-plane that ends at the origin will no longer be exponentially 
suppressed. However, our assumption that $f(s)$ has a finite limit as 
$s\rightarrow 0$ implies that if $f\sim s^\kappa$ then
$\kappa>0$ and this behaviour is power law suppressed. Also the expression
for $I_\lambda^N$ simplifies to $\sum_{n=0..N}a_{n}\lambda^{n}/n!$.
In our example $I_\lambda=e^{-\lambda}/\lambda-1/\lambda$
and the second term demonstrates this power law suppression.
Again $I_\lambda^{19}$ and $I_\lambda^{18}$ agree for
$\lambda$ up to about $7$ yielding an approximation of 
$-0.1416$ to $f(0)=0$. 
\FIGURE{\centerline{\epsfysize=9cm\epsfbox{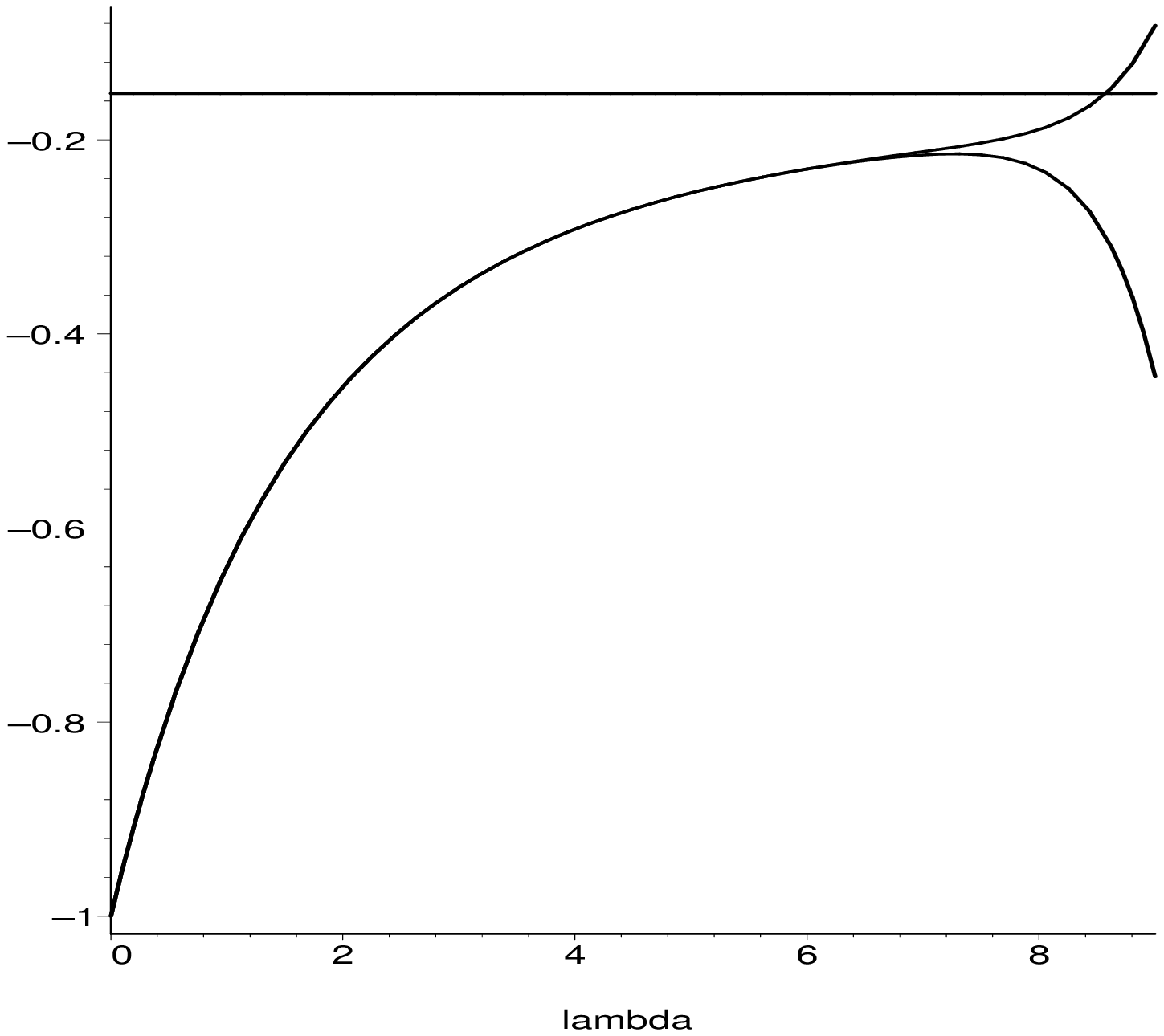}\caption[series]{\label{resumex2} $I_\lambda^{19}$ (upper curve) and $I_\lambda^{18}$
for $s\rightarrow 0$}}}

This approximation can be greatly improved 
by changing the complex variable $s$, for
example by considering $f(s^{a})$ with $a>1$ instead of $f(s)$,
so that now
\be
I_\lambda^N=\sum_{n=0..N}a_{n}\lambda^{na}/\Gamma(na+1)
\label{I}
\ee
Increasing 
$a$ reduces the size of the last term in $I_{\lambda}^{N}$
in comparison to the penultimate term, because of the $\Gamma$ 
function
in (\ref{I}), enabling us to take a larger value of $\lambda$,
however increasing $a$ enlarges the region of 
non-analyticity. Initially this will result in singularities 
occurring at values of $s$ with large negative parts, causing 
oscillations
in the plot of $I_{\lambda}^{N}$ against $\lambda$. This can be 
seen in our example, in Figure \ref{resumex3}.

\FIGURE{\centerline{\epsfysize=9cm\epsfbox{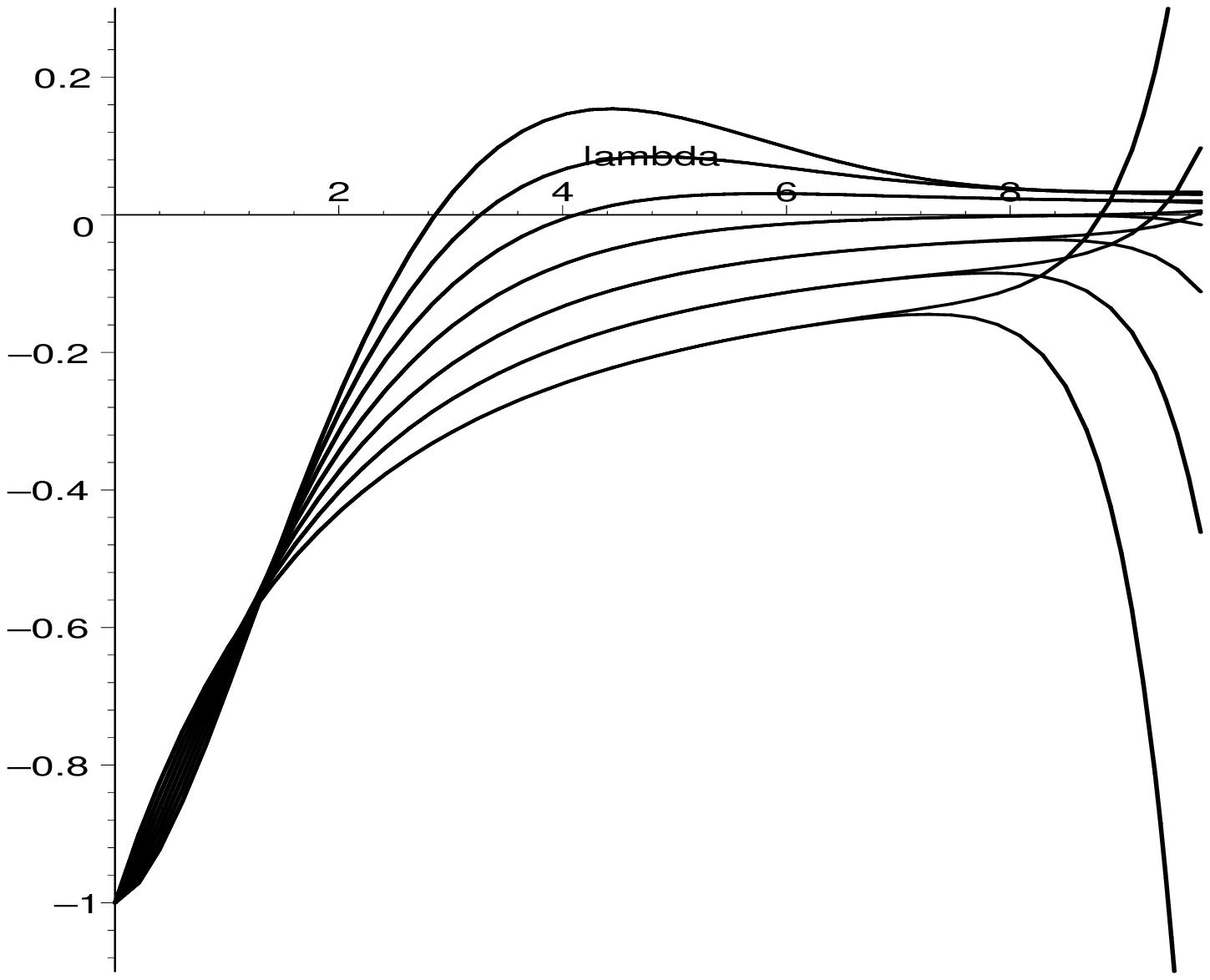}\caption[series]{\label{resumex3}$I_{\lambda}^{19}$ and $I_{\lambda}^{18}$
for 
$a=1$ (lowest curve), $1.1,..,1.6$ }}}

Maxima appear for $a=1.4$, $1.5$ and $1.6$ due to these 
oscillations. Increasing $a$ still further can cause the singularities
to migrate to the right half-plane, spoiling our 
construction altogether. The position of these dominant singularities 
can be estimated by studying the shape of the curve, 
thus the $\exp(-\lambda)$ decay in our example for $a=1$
can be seen by fitting numerically the curve of $I_{\lambda}^{N}$ 
to an exponential. To improve our approximation 
we need to increase $a$ as much as we can without encountering
these pathologies. We can do this by taking $a$ 
as large as possible whilst maintaining $I_{\lambda}^{N}$
as a monotonic function of $\lambda$ in the region where it
approximates $f(0)$, $\lambda<\lambda*$, (we may have to exclude a 
region
close to the origin if $I_{\lambda}$ has a turning point
not due to a singularity occurring at complex $s$, this
complication can be resolved by studying the $N$-dependence). So in our example we would take 
$a=1.348004$ and obtain the approximation $0.013$ when we 
require that $I_{\lambda}^{18}$ and $I_{\lambda}^{19}$
differ by $0.01$ at $\lambda*=9.77$.

The precision with which we require $I_{\lambda}^{19}$ and $I_{\lambda}^{18}$
to agree introduces arbitrariness into $\lambda*$, however because we tune
$a$ so that $I_{\lambda}^{19}$ is quite flat for $\lambda\approx \lambda*$
the estimate we obtain is fairly insensitive to this choice of precision,
so for example if we require that $I_{\lambda}^{19}$ and $I_{\lambda}^{18}$
differ by only $0.001$ (instead of $0.01$) at $\lambda*$ the estimate changes to $0.0115$
even though $\lambda*$ has changed to $8.93$.
Note that $I_{\lambda}^{18}$ is considerably less flat than
$I_{\lambda}^{19}$ in the vicinity of $\lambda*$. It has a single maximum
in this region, whereas the flat portion of the latter turns into
two turning points that are close together if $a$ is slightly increased.
To minimise arbitrariness we should in general
base our estimates on the flatter curve, which means using   
$I_{\lambda}^{N}$ with $N$ odd in this example.
\FIGURE{\centerline{\epsfysize=9cm\epsfbox{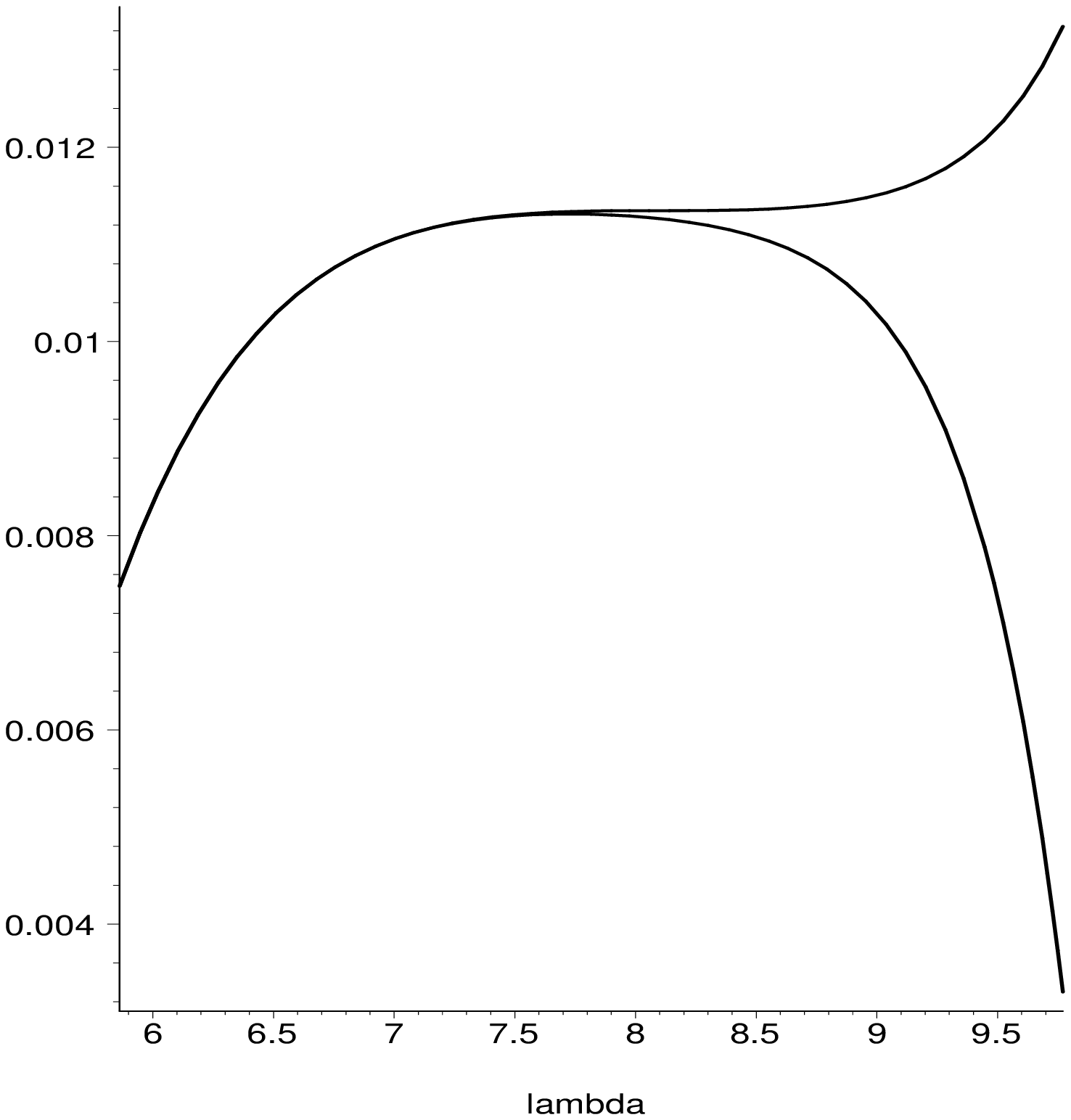}\caption[series]{\label{resumex8}$I_{\lambda}^{19}$ (upper
curve) and $I_{\lambda}^{18}$
for 
$a=1.348004$ in the vicinity of $\lambda*$}}}

The integral $I_{\lambda}$ is the Borel transform of
the function $f$ centred at $s_{0}$. Our approach based
on studying the large-$\lambda$ behaviour to reconstruct
the original function has the advantage over the usual method of avoiding 
the need to
analytically continue the transform, (for example
using Pad\'e approximants), prior to inversion using the Laplace
transform.

In our application to the \Sc equation $s$ is a sharp momentum
cut-off. It has been observed that in the different, but related context,
of Polchinski's flow equation that a smooth cut-off increases the rate of convergence
of a series solution. The reverse occurs here.
Suppose that the series $A_N$ resulted from an
integral with a sharp cut-off as might be the case for example for a one-loop
Feynman diagram. Thus
\be
A_N=\sum_{n=0}^N a_n/s^n=\int_0^{1/s} dk\,\sum_{n=0}^N a_n\,nk^{n-1}
\ee
Our transform $I^N_\lambda$ is $\sum_{n=0..N}a_{n}\lambda^{n}/n!$
as before, but if we replace the sharp cut-off by a Gaussian
one we obtain an alternative to $A_N$
\be
\int_0^{\infty} dk\,\sum_{n=0}^N a_n\,nk^{n-1}\,e^{-s^2k^2}
={1\over 2}\sum_{n=0}^N \Gamma(n/2)\,a_n/s^n 
\ee
with a transform $\sum_{n=0..N}\Gamma(n/2)\,a_{n}\lambda^{n}/(2n!)$
which clearly converges much more slowly (if at all) due to the
$\Gamma(n/2)$ in the  numerator. That this is the reverse to what 
occurs in the Polchinski flow equation in \cite{mo} is  
because in that application the corresponding expansions 
involve positive powers of $s$ rather than negative ones
so that $\Gamma(n/2)$ appears instead in the denominator 
and enhances convergence.

\vfill
\eject

\section{Surface Critical Scaling in The Ising Model}

$\p^{4}$ theory is the Landau-Ginzburg theory that describes the
critical behaviour of the Ising model \cite{Itzyk}, \cite{Par}. The vacuum wave-functional
has a functional integral representation as a Euclidean field
theory with a boundary corresponding to the quantisation
surface
\bq
&&\Psi [\p]=\int {\cal D\phi}\,\exp\Big(-\int_{-\infty}^{0} dt
\int_{-\infty}^{\infty} dx \left( {1\over 2}\left (\dot\phi^{2}+
\phi^{\prime 2}
+m_{0}^{2}\,\phi^{2}\right)+{g_{0}\over 4!}\,\phi^{4}\right)\nonumber\\
&&\quad\quad\quad\quad\quad\quad\quad\quad\quad\quad\quad\quad\quad\quad\quad
+\int_{-\infty}^{\infty} dx \varphi_{0}\,\dot\phi(x,0)\Big)
\label{fint}
\eq
where the $\phi$ satisfies Dirichlet boundary conditions
$\phi(x,0)=0$, and a regulator is imposed
(which we take as the same momentum cut-off as before). 
$\p_{0}$ appears as a source term
coupled to $\dot\phi$ so that functional differentiation
results in an insertion of $\dot\phi$.
For a particular value of $g$ the $\mu\rightarrow 0$ limit of 
$\Psi$ models the critical behaviour of the 
two-dimensional Ising model with a boundary on which the spins
$\sigma\sim\phi$ obey Dirichlet boundary conditions,
this is the Ordinary Transition. 
The correlator $\langle \dot\sigma(0,0)\,\dot\sigma(x,0)\rangle$
can be computed from the exact Onsager solution or from the
conformal field theory four-point function using the method
of images and shown to depend on $|x|$ as $1/|x|$, \cite{Cardy}.
So in the Conformal Field Theory limit of $\p^{4}$
that part of $W$ quadratic in the field,$\int dx\,dy 
\,\p_0(x)\,\p_0(y)\,\Gamma_0(x-y)=\int dx\,dy 
\,\p(x)\,\p(y)\,\Gamma(x-y)$, should have
$\Gamma\propto1/|x-y|$.

We have studied massive $\p^{4}$ theory, but the Renormalisation Group 
provides a means of extracting the massless behaviour \cite{Itzyk}, \cite{Par}. 
It starts from 
the observation that at short distances the functional integral
(\ref{fint}) expressed as it is in terms of bare quantities
is approximately independent of $m_{0}$. This is based on a 
perturbative analysis, because in terms of Feynman diagrams
differentiating $\Psi$ with respect to $m_{0}$ introduces a 
$\phi^{2}$ insertion in the bulk involving two propagators
that are suppressed at large momentum. Now, from the functional integral
representation $\Gamma_0$ is a function of the bare variables
$g_0$, $m_0$ and the cut-off and is related by wave-function 
renormalisation to $\Gamma$ which is given by our solution in terms  of $\mu$ and $g$. By dimensional analysis $Z$ depends on
$\mu$ and $s$ only in the combination $\mu s$ whilst
$\Gamma$ depends on $x-y$ as $F(\mu(x-y),g)/|x-y|^2$
with $F$ a dimensionless function,
so
\be
\Gamma_0(x-y,g_0,m_0,s)=Z^{-1}(\mu s,g)\,F(\mu(x-y),g)/|x-y|^2
\ee
Since $\Gamma_0$ can be taken to be independent of $m_0$
for small $|x-y|$ the left-hand-side does not change if
$m_0$ is varied whilst $g_0$ and $s$ are held fixed.
As $m_0$ varies $\mu$ and $g$ will vary together, so that
$g$ can be thought of as a function of $\mu$ with a $\beta$-function, $\beta=dg/d\log\mu=\beta(g)$. Since $\Gamma_0$ does not
change if we vary $\mu$ and $g$ from $\mu_1$ and $g_1$
to $\mu_2$ and $g_2=g(\mu_1,\mu_2,g_1)$we obtain
\be
F(\mu_2(x-y),g_2)={Z(\mu_2 s,g(\mu_1,\mu_2,g_1))\over Z(\mu_1 s,g_1)}F(\mu_1(x-y),g_1)\label{F}
\ee
Now $\gamma\equiv d\log Z/d\log \mu$ is a function of $g$,
so
\be
{Z(\mu_2 s,g(\mu_1,\mu_2,g_1))\over Z(\mu_1 s,g_1)}
=
\exp\int_{\mu_1}^{\mu_2}{d\mu\over\mu}\gamma
=\exp\int_{g_2}^{g_1}{dg\over\beta}\gamma
\label{zed}
\ee
If $g_*$ is a zero of the $\beta$-function and if we assume that
$\gamma$ is continuous then for $g_1$, $g_2$ close to $g_*$
(\ref{zed}) behaves like
\be
\exp\int_{g_2}^{g_1}{dg\over (g-g_{*})\,\beta'(g_*)}\gamma(g_*)
\approx \left({\mu_2\over\mu_1}\right)^{\gamma(g_*)}
\ee
If in addition we assume that $F(x,g)$ is continuous near
$g_*$ then the limit as $g_1\rightarrow g_*$ of 
(\ref{F}) is 
\be
F(\mu_2(x-y),g_*)=\left({\mu_2\over\mu_1}\right)^{\gamma(g_*)}
F(\mu_1(x-y),g_*)
\ee
which implies the homogeneous dependence on $x-y$:
$F\sim |x-y|^{\gamma(g_*)}$
that is characteristic of a Conformal Field Theory.

The correspondence betwen the Ising model and 
$\p^4$ theory requires that we should be able to obtain 
$\gamma(g_*)=1$ from our solution. The series
given in the Appendix express $g_0$ and $Z$ as functions
of $\mu$, $s$ and $g$. Differentiating these with respect
to $\log \mu$ whilst keeping $g_0$ and $s$ fixed gives
\be
0={\partial g_0\over\partial \log \mu}+
\beta\, {\partial g_0\over\partial g},
\quad \gamma\,Z={\partial Z\over\partial \log \mu}
+\beta \,{\partial Z\over\partial g}\,.
\ee
From which $\beta$ and $\gamma$ are obtained as power 
series in $1/s$ with the results also given in the Appendix 
up to order $1/s^{17}$. We have calculated these functions for 
a variety of values of $g$
using the transformation of 
Section 3 to extract the $s\rightarrow 0$ limit.
The results are shown in Figure \ref{resumex5} and demonstrate that the
$\beta$-function has a zero at $g$ roughly equal to
$2$ where $\gamma$ is approximately unity in agreement
with the Ising model result. 
Studying the region
in the vicinity of $g=2$ leads to values of $1.955$
for the zero and $0.961$ for the corresponding value of $\gamma$.
\FIGURE{\centerline{\epsfysize=9cm\epsfbox{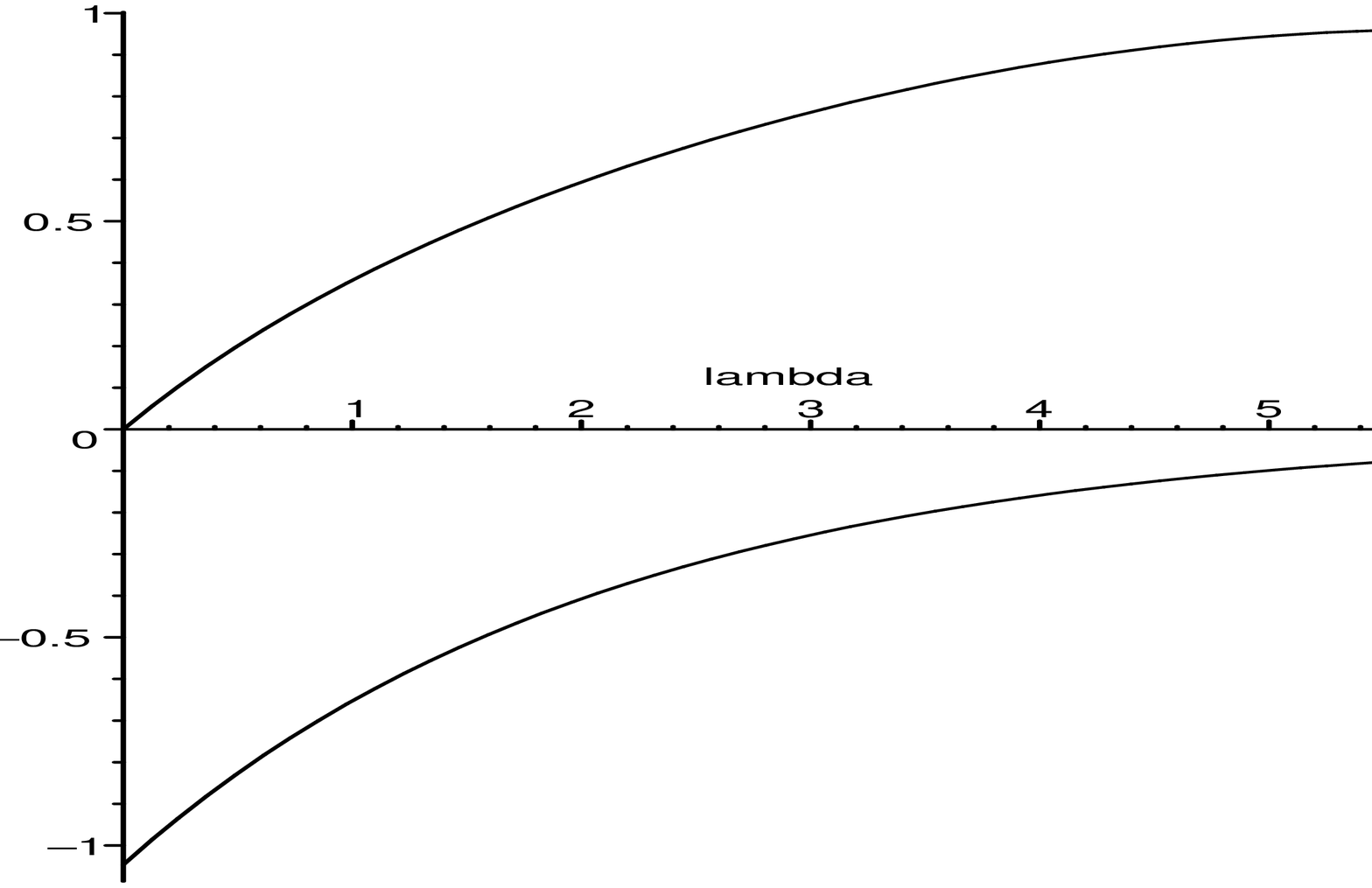}\caption[series]{\label{resumex6}$I_{\lambda}^{17}$ (upper curve) and
$\exp(m\lambda+c)/m$ approximating the error.}}}

Comparison with the exact result shows a departure from it by four per cent,
but we can estimate the error {\it a priori} by examining the shape of 
$I_{\lambda}^{N}$. For small $\lambda$ the error is mainly composed of
exponential terms that are suppressed as $\lambda$ increases,
we can estimate the dominant term
by fitting $dI_{\lambda}^{N}/d\lambda$ to an expression of the form
$\exp(m\lambda+c)$ for $0<\lambda<0.9\lambda *$
(the region is only taken up to $0.9\lambda*$ rather 
than $\lambda$ because the $dI_{\lambda}^{17}/d\lambda$
and $dI_{\lambda}^{16}/d\lambda$ diverge for smaller $\lambda$). 
In Figure \ref{resumex6} we plot $I_{\lambda}^{17}$
and $\exp(m\lambda+c)/m$ which shows that the exponential terms account for about 
six per cent of the value of $I_{\lambda}^{N}$, this is of course only a rough estimate.

This approximation of the anomalous dimension depends on the number of terms
we keep in the series expansions, which in turn determines
the type of terms we need to retain in our
local expansions of $W$ and $U$. So far we have 
worked to order $1/s^{n}$ with $n=17$ which has required including
terms with $F$ fields and $D$ derivatives such that $F+2D\le 40$.
Since the true continuum limit should be obtained as $n\rightarrow\infty$
it is worthwhile to see how the results vary with $n$.
\FIGURE{\centerline{\epsfysize=9cm\epsfbox{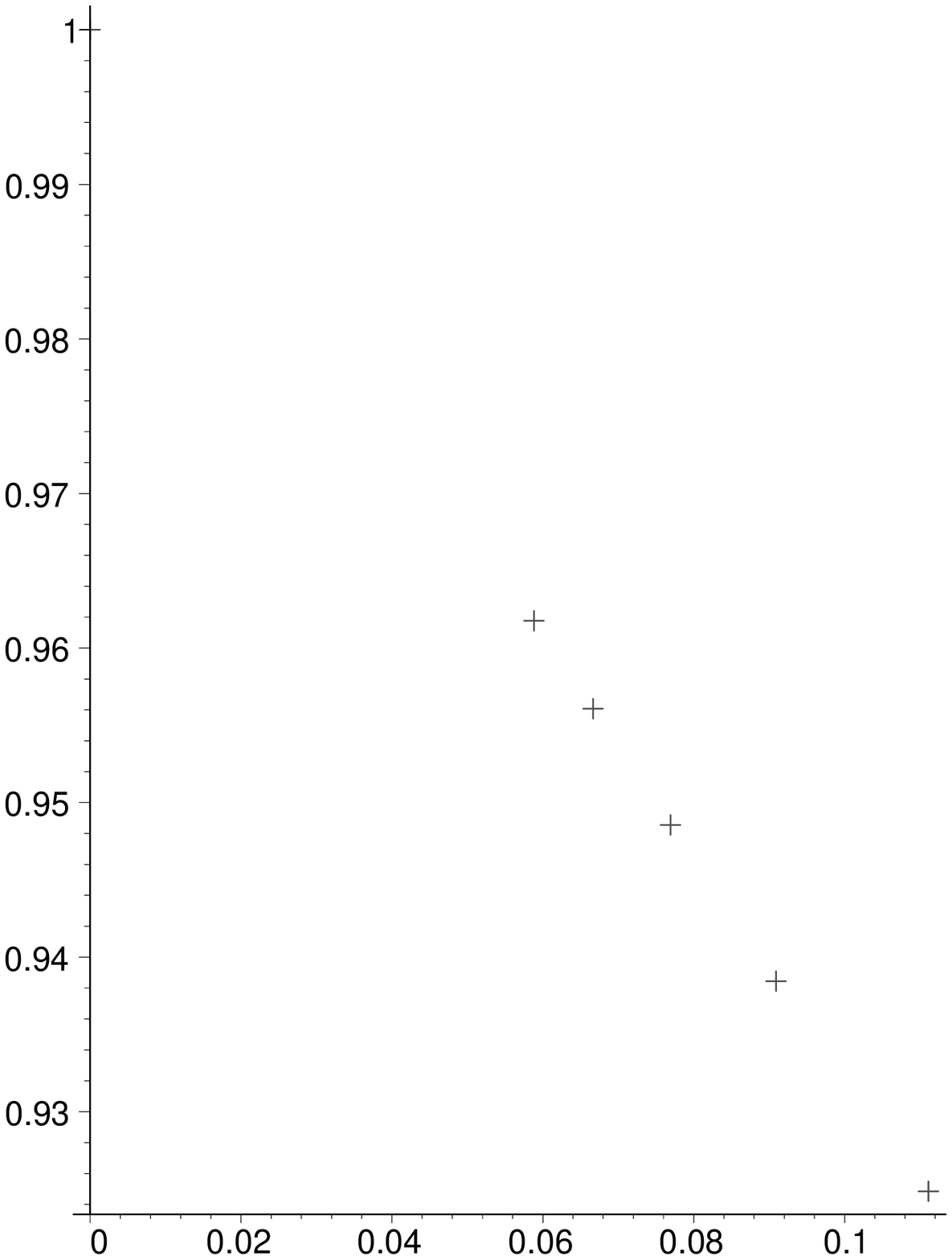}\caption[series]{\label{resumex4}The approximation to $\eta$ for $n=17,\,15,\,13,\,11,\,9$}}}
In Figure \ref{resumex4} we
plot the approximations obtained by keeping $n$ terms against $1/n$
for $n=17,\,15,\,13,\,11,\,9$. We also show the exact result 
as though it would occur at $n=\infty$, to guide the eye.
The plot shows roughly linear behaviour, and when a 
straight-line fit through these results is extrapolated
we obtain a value of $1.003$ for the anomalous dimension,
which is in very good agreement with the exact Ising Model result.

We can compare this result with the predictions of one-loop
perturbation theory. The vacuum functional was computed in the
semi-classical approximation in \cite{Marcos} (the perturbative
treatment of the three-dimensional case is given in \cite{Shp})
where
the coefficients of $\p_0^4$ and $\p_0^{\prime 2}$ 
in $W$ were found to be $-g_0/(96 M)+3.973.10^{-4}g_0^3/M^3$
and $-1/(4M)+g_0/(48\pi M)$, respectively, where $M$ is related
to the particle mass by $M=\mu-g_0/(8\pi \mu)$. These imply
that at one-loop order $\beta=-2g+0.617g^2$ and $\gamma=g/\pi$,
so that $g_*\approx 3.242$ and $\gamma(g_*)\approx 1.032$,
which is slightly more accurate than our estimate based on series with
17 terms, but considerably less accurate than the extrapolated
value of $1.003$. 

The expansions in the appendix can be re-ordered to yield estimates
for the coefficients of the loop expansion. For example, if we
retain only those terms up to and including second order in $g$ in 
the beta-function we obtain

\bq
\beta\approx &&-2g+
g^2\Big({\frac {15}{8}}\,{\frac {1}{\pi \,\mu\,s}}-{\frac {65}{72}}\,{\frac {1
}{{\mu}^{3}{s}^{3}\pi }}+{\frac {3773}{5760}}\,{\frac {1}{{\mu}^{5}{s}
^{5}\pi }}-{\frac {179}{336}}\,{\frac {1}{{\mu}^{7}{s}^{7}\pi }}
\nonumber\\
&&+{\frac {341759}{746496}}\,{\frac {1}{{\mu}^{9}{s}^{9}\pi }}-{\frac {
1112423}{2737152}}\,{\frac {1}{{\mu}^{11}{s}^{11}\pi }}+{\frac {
38151475}{103514112}}\,{\frac {1}{{\mu}^{13}{s}^{13}\pi }}
\nonumber\\
&&-{\frac {182364253}{537477120}}\,{\frac {1}{{\mu}^{15}{s}^{15}\pi }}+
{\frac {9235090201}{29238755328}}\,{\frac {1}{{\mu}^{17}{s}^{17}\pi }}\Big)
\eq
Using the Borel resummation technique to find the $s\rightarrow 0$ limit of 
the coefficient of $g^2$ gives $0.633$ which is in reasonable agreement with the
one-loop value $0.617$ given that this series has so few terms in comparison with
the previous series we have considered. Similarly the one-loop contribution to $\gamma$
is approximated by
\bq
\gamma\approx &&g\Big({\frac {9}{8}}\,{\frac {1}{\pi \,\mu\,s}}-5/8\,{\frac {1}{{\mu}^{3}{s}
^{3}\pi }}+{\frac {63}{128}}\,{\frac {1}{{\mu}^{5}{s}^{5}\pi }}-{
\frac {27}{64}}\,{\frac {1}{{\mu}^{7}{s}^{7}\pi }}
\nonumber\\
&&
{\frac {385}{1024}}\,{\frac {1}{{\mu}^{9}{s}^{9}\pi }}-{\frac {351}{
1024}}\,{\frac {1}{{\mu}^{11}{s}^{11}\pi }}+{\frac {10395}{32768}}\,{
\frac {1}{{\mu}^{13}{s}^{13}\pi }}
\nonumber\\
&&
-{\frac {2431}{8192}}\,{\frac {1}{{\mu}^{15}{s}^{15}\pi }}+{\frac {73359}{262144}}\,{\frac {1}{{\mu}^{17}{s}^{17}\pi }}\Big)
\eq
Resummation yields the value $0.336 g$ which should be compared with
$g/\pi\approx 0.318 \,g$. 

\vfill
\eject
We can go further and estimate the two-loop contribution to the beta-function
from the series
\bq
&&g^3\Big(-9/2\,{\frac {1}{\pi \,{\mu}^{2}{s}^{2}}}+{\frac {65}{18}}\,{\frac {1}
{\pi \,{\mu}^{4}{s}^{4}}}+{\frac {5975}{10368}}\,{\frac {1}{\pi \,{\mu
}^{6}{s}^{6}}}-{\frac {1082059}{60480}}\,{\frac {1}{\pi \,{\mu}^{8}{s}
^{8}}}
\nonumber\\
&&
{\frac {637469833}{7464960}}\,{\frac {1}{\pi \,{\mu}^{10}{s}^{10}}}-{
\frac {73790276321}{215550720}}\,{\frac {1}{\pi \,{\mu}^{12}{s}^{12}}}
+{\frac {52328568942469}{39852933120}}\,{\frac {1}{\pi \,{\mu}^{14}{s}
^{14}}}
\nonumber\\
&&
{\frac {1607134369983293}{322808758272}}\,{\frac {1}{\pi \,{\mu}^{16}{s}^{16}}}
\Big)
\eq
which gives $-0.813g^2$ on resummation, so to two-loop order the $\beta$-function
ceases to have a zero. The disappearance (and reappearance) of the zero in successive orders
of perturbation theory is encountered elsewhere and requires a skilful handling
of the perturbation expansion, for example with the use of Pad\'e-Borel resummation in
$g$, \cite{Shp}.
This is to be contrasted with the results from the \Sc equation for which
there is always a zero in the  $\beta$-function although its
position is corrected at each increasing order of the approximation.

\section{Conclusions}
We have solved the functional \Sc equation for the 
vacuum of $\p^4$ theory 
and for a state describing a stationary particle by expanding these states in terms of local functionals. The solutions were obtained as power series in $1/s$, where 
$s$ is a short-distance cut-off. The $s\rightarrow 0$ limits of these series were 
found using a variant of Borel re-summation that avoids the use of
analytic continuation. A standard Renormalisation Group argument was 
used to compute an anomalous scaling dimension in the
massless theory. Extrapolating in the number of terms of the series
gave the result 1.003 for this scaling dimension which agrees well
with the exact result of unity obtained from the conformal field theory
treatment of the surface critical scaling
of the Ising model at the Ordinary Transition.

We have only developed the solution to the point where we could make the comparison 
with the Ising model. Further work remains to be done to 
describe the particle dynamics in this approach,
for example by checking Lorentz invariance and constructing 
an S-matrix. Neither have we computed inner products of states, although their form 
is determined by the Hermiticity of the Hamiltonian and they should
also be calculable as re-summed expansions in $1/s$.

We concentrated on two dimensional scalar theory as this allowed the
series expansions to be obtained using only modest computing resources.
Similar resources should be sufficient to study other models such as 
Landau-Ginzburg theories with higher order polynomial interactions
and integrable models with exponential interactions where the wealth of exact
results would provide useful tests of the method. There is nothing intrinsic
to the approach that limits it to two dimensions, 
so it should be possible to apply it to critical
phenomena in 3 dimensions and gauge theories in three and four.
Indeed the method was motivated by the desire to systematise
a heuristic model of the Yang-Mills vacuum that gave an accurate formula for
the string tension in three dimensions. Applying it to
that case to compute series with large numbers of terms 
would require some effort, so we cannot at the moment 
explain why the formula for the string-tension 
agrees accurately with Monte-Carlo results, despite being
based on just the leading term of the local expansion.
However we can understand why the corresponding calculation
in four-dimensions appears to give a string-tension that diverges with the
cut-off: this is just the first term in a power series that should 
re-sum to a finite result. As in the case of scalar field theory
standard perturbation theory is obtained by organising the 
local expansion in powers of the coupling. In this context
it has already been shown how the
usual one-loop beta-function for Yang-Mills theory is obtained
in this approach \cite{Marcos2}.

\vfill\eject
\appendix
\section{Solutions}

The solutions listed below for the series expansions of $Z$, $g_0$, $\beta$ and $\gamma$
were computed up to order $1/s^{17}$ using MAPLE, working with rational numbers. The exact
values for the cofficients have been evaluated to ten significant figures to save space.

Solving the \Sc equation yields the series expansion of the wave-function renormalisation, $Z=\sum_{n}z_n/(\mu s)^n$,
with the first 18 coefficients given by 

$$z_{{0}}= 1.0
$$ $$z_{{1}}=- 0.1193662073\,g
$$ $$z_{{2}}= 0.05066059180\,{g}^{2}
$$ $$z_{{3}}= 0.03978873576\,g- 0.07596748147\,{g}^{3}
$$ $$z_{{4}}= 0.1658372186\,{g}^{4}- 0.04239306467\,{g}^{2}
$$ $$z_{{5}}= 0.1322116938\,{g}^{3}- 0.4702098769\,{g}^{5}- 0.02238116387\,
g
$$ $$z_{{6}}= 1.600512277\,{g}^{6}- 0.4288302602\,{g}^{4}+ 0.02320184743\,{
g}^{2}
$$ $$z_{{7}}= 0.01492077591\,g+ 1.643970766\,{g}^{5}- 6.280656209\,{g}^{7}-
 0.2150394698\,{g}^{3}
$$ $$z_{{8}}=- 7.118564002\,{g}^{6}+ 27.72021990\,{g}^{8}+ 0.03759400389\,{
g}^{2}+ 0.9449811249\,{g}^{4}
$$ $$z_{{9}}=- 0.01087973244\,g- 4.714821300\,{g}^{5}+ 34.16960776\,{g}^{7}
$$ $$- 135.3903272\,{g}^{9}+ 0.3954218093\,{g}^{3}
$$ $$z_{{10}}= 723.5185086\,{g}^{10}+ 25.14159983\,{g}^{6}- 179.4495800\,{g
}^{8}$$ $$- 0.2452396545\,{g}^{2}- 2.090535257\,{g}^{4}
$$ $$z_{{11}}= 0.008392936450\,g- 4195.152125\,{g}^{11}+ 14.06398176\,{g}^{
5}$$ $$- 143.8822880\,{g}^{7}+ 1021.740867\,{g}^{9}- 0.9435065158\,{g}^{3}
$$ $$z_{{12}}= 26222.36179\,{g}^{12}- 6265.013026\,{g}^{10}- 92.19320030\,{
g}^{6}$$ $$+ 880.6482723\,{g}^{8}+ 0.9593860756\,{g}^{2}+ 3.790793362\,{g}^
{4}
$$ $$z_{{13}}=- 0.006731834444\,g- 175775.0461\,{g}^{13}+ 41156.19353\,{g}^
{11}$$ $$- 45.34168199\,{g}^{5}+ 626.1457656\,{g}^{7}- 5746.450353\,{g}^{9}
$$ $$+ 3.174250642\,{g}^{3}
$$ $$z_{{14}}=- 288442.7166\,{g}^{12}+ 1258090.875\,{g}^{14}+ 39859.60610\,
{g}^{10}$$ $$+ 375.4551915\,{g}^{6}- 4438.133512\,{g}^{8}- 3.420232517\,{g}
^{2}$$ $$+ 9.605144838\,{g}^{4}
$$ $$z_{{15}}= 0.005556434780\,g+ 2149122.481\,{g}^{13}- 293133.1749\,{g}^{
11}$$ $$+ 138.6466218\,{g}^{5}- 3087.933824\,{g}^{7}+ 32967.21250\,{g}^{9}$$ $$-
 14.23407683\,{g}^{3}- 9578735.713\,{g}^{15}
$$ $$z_{{16}}= 2280136.160\,{g}^{12}- 16970850.80\,{g}^{14}- 256999.1253\,{
g}^{10}$$ $$- 1543.206840\,{g}^{6}+ 25586.11980\,{g}^{8}+ 11.94349655\,{g}^
{2}$$ $$- 272.9892577\,{g}^{4}+ 77323626.71\,{g}^{16}
$$ $$z_{{17}}=- 0.004688241844\,g+ 75.31011486\,{g}^{3}+ 16703.89651\,{g}^{7}
$$ $$-141.2669974\,{g}^{5}+ 141642086.6\,{g}^{15}- 18718142.21\,{g}^{13}$$
$$+2102889.466\,{g}^{11}- 217344.7165\,{g}^{9}- 659842155.7\,{g}^{17}$$

\vfill\eject
Solving the \Sc equation yields the series expansion of the bare-coupling, 
$g_0=\mu^2\sum_{n}\kappa_n/(\mu s)^n$,
with the first 18 coefficients given by

$$\kappa_{{0}}= 1.0\,g
$$ $$\kappa_{{1}}= 0.1989436788\,{g}^{2}
$$ $$\kappa_{{2}}=- 0.03641230036\,{g}^{3}
$$ $$\kappa_{{3}}= 0.08207763547\,{g}^{4}- 0.05747261834\,{g}^{2}
$$ $$\kappa_{{4}}=- 0.1813715346\,{g}^{5}+ 0.02286762825\,{g}^{3}
$$ $$\kappa_{{5}}= 0.5276961593\,{g}^{6}- 0.1344338864\,{g}^{4}+ 0.02978628969\,
{g}^{2}
$$ $$\kappa_{{6}}=- 1.827461455\,{g}^{7}+ 0.4342310123\,{g}^{5}+ 0.02099375315\,
{g}^{3}
$$ $$\kappa_{{7}}= 0.2040031164\,{g}^{4}- 0.01884175582\,{g}^{2}- 1.725204731\,{
g}^{6}+ 7.268618386\,{g}^{8}
$$ $$\kappa_{{8}}=- 0.8376851525\,{g}^{5}- 0.1619837103\,{g}^{3}+ 7.650541972\,{
g}^{7}- 32.43432692\,{g}^{9}
$$ $$\kappa_{{9}}=- 0.3373967928\,{g}^{4}+ 0.01324798773\,{g}^{2}+ 4.478764811\,
{g}^{6}$$ $$- 37.43268679\,{g}^{8}+ 159.8791617\,{g}^{10}
$$ $$\kappa_{{10}}= 1.190013912\,{g}^{5}+ 0.6263468208\,{g}^{3}- 24.78826072\,{g
}^{7}$$ $$+ 199.6885243\,{g}^{9}- 861.1578312\,{g}^{11}
$$ $$\kappa_{{11}}= 0.7212680080\,{g}^{4}- 0.009951254171\,{g}^{2}- 10.64530019
\,{g}^{6}$$ $$+ 146.1049618\,{g}^{8}- 1152.010150\,{g}^{10}+ 5027.712587\,{
g}^{12}
$$ $$\kappa_{{12}}= 3.152841368\,{g}^{5}- 2.174458226\,{g}^{3}+ 76.02682717\,{g}
^{7}$$ $$- 915.5371834\,{g}^{9}+ 7143.555824\,{g}^{11}- 31617.81483\,{g}^{
13}
$$ $$\kappa_{{13}}=- 2.418827270\,{g}^{4}+ 0.007821150456\,{g}^{2}+ 13.86178637
\,{g}^{6}$$ $$- 549.6708328\,{g}^{8}+ 6091.031330\,{g}^{10}- 47386.89418\,{
g}^{12}$$ $$+ 213090.6078\,{g}^{14}
$$ $$\kappa_{{14}}=- 68.20413916\,{g}^{5}+ 7.396453034\,{g}^{3}- 178.3373919\,{g
}^{7}$$ $$+ 4076.316308\,{g}^{9}- 42945.16713\,{g}^{11}+ 334961.5945\,{g}^{
13}$$ $$- 1532576.001\,{g}^{15}
$$ $$\kappa_{{15}}= 11.95109263\,{g}^{4}- 0.006353030446\,{g}^{2}+ 151.2121465\,
{g}^{6}$$ $$+ 1867.345135\,{g}^{8}- 31358.18697\,{g}^{10}+ 320273.7334\,{g}
^{12}$$ $$- 2514689.736\,{g}^{14}+ 11719597.0\,{g}^{16}
$$ $$\kappa_{{16}}= 819.5644108\,{g}^{5}- 25.22638048\,{g}^{3}- 790.3286415\,{g}
^{7}$$ $$- 17670.92565\,{g}^{9}+ 251437.2784\,{g}^{11}- 2521699.943\,{g}^{
13}$$ $$+ 19992199.42\,{g}^{15}- 94979792.35\,{g}^{17}
$$ $$\kappa_{{17}}=- 167874034.0\,{g}^{16}- 2682.130056\,{g}^{6}+ 0.005291499834\,{g}^{2}
$$ $$- 69.91144120\,{g}^{4}+ 436.3776317\,{g}^{8}+ 163982.6175\,{g}^{10}-
$$ $$2105884.503\,{g}^{12}+ 20923210.23\,{g}^{14}+ 813423914.7\,{g}^{18}$$

\vfill\eject

The solutions for $Z$ and $g_0$ lead to the series expansion of the $\beta$-function,
$\beta=\sum_{n}\beta_n/(\mu s)^n$, with the first 18 coefficients given by 

$$\beta_{{0}}=- 2.0\,g
$$ $$\beta_{{1}}= 0.5968310364\,{g}^{2}
$$ $$\beta_{{2}}=- 0.4559453262\,{g}^{3}
$$ $$\beta_{{3}}= 0.9853095730\,{g}^{4}- 0.2873630916\,{g}^{2}
$$ $$\beta_{{4}}= 0.3658820519\,{g}^{3}- 2.814252613\,{g}^{5}
$$ $$\beta_{{5}}= 0.2085040278\,{g}^{2}+ 9.833762561\,{g}^{6}- 1.749096318\,{g}
^{4}
$$ $$\beta_{{6}}= 0.05839063195\,{g}^{3}- 39.74109388\,{g}^{7}+ 7.374962709\,{g
}^{5}
$$ $$\beta_{{7}}=- 0.1695758024\,{g}^{2}+ 180.4235628\,{g}^{8}- 34.76674176\,{g
}^{6}+ 2.702926953\,{g}^{4}
$$ $$\beta_{{8}}=- 903.9912677\,{g}^{9}- 1.812756260\,{g}^{3}+ 178.4867048\,{g}
^{7}- 15.43817775\,{g}^{5}
$$ $$\beta_{{9}}= 0.1457278651\,{g}^{2}- 990.1865501\,{g}^{8}+ 96.47016296\,{g}
^{6}$$ $$- 4.096644239\,{g}^{4}+ 4940.346009\,{g}^{10}
$$ $$\beta_{{10}}= 5898.620256\,{g}^{9}- 29206.19354\,{g}^{11}+ 8.652316689\,{g
}^{3}$$ $$- 615.2794259\,{g}^{7}+ 24.26339449\,{g}^{5}
$$ $$\beta_{{11}}=- 0.1293663043\,{g}^{2}+ 4091.802192\,{g}^{8}- 241.2179221\,{
g}^{6}$$ $$+ 7.386237270\,{g}^{4}- 37554.52475\,{g}^{10}+ 185620.8411\,{g}^
{12}
$$ $$\beta_{{12}}=- 28511.70908\,{g}^{9}+ 254595.6560\,{g}^{11}- 34.68565605\,{
g}^{3}$$ $$+ 1992.597606\,{g}^{7}+ 59.07548891\,{g}^{5}- 1262105.326\,{g}^{
13}
$$ $$\beta_{{13}}= 9144064.399\,{g}^{14}+ 0.1173172568\,{g}^{2}- 16205.08595\,{
g}^{8}$$ $$+ 289.2141186\,{g}^{6}- 26.54957599\,{g}^{4}+ 208570.6632\,{g}^{
10}$$ $$- 1832311.357\,{g}^{12}
$$ $$\beta_{{14}}= 133241.9688\,{g}^{9}- 1602568.994\,{g}^{11}+ 133.0389542\,{g
}^{3}$$ $$- 4843.070749\,{g}^{7}- 1489.424050\,{g}^{5}+ 13962796.56\,{g}^{
13}$$ $$- 70350835.71\,{g}^{15}
$$ $$\beta_{{15}}=- 112396967.2\,{g}^{14}- 0.1080015175\,{g}^{2}+ 57364.96045\,
{g}^{8}$$ $$+ 4603.125796\,{g}^{6}+ 573028054.8\,{g}^{16}+ 177.9010353\,{g}
^{4}$$ $$- 1123852.943\,{g}^{10}+ 12930212.94\,{g}^{12}
$$ $$\beta_{{16}}=- 602711.1263\,{g}^{9}+ 9794809.468\,{g}^{11}- 504.4372323\,{
g}^{3}$$ $$- 24799.33347\,{g}^{7}+ 19590.50509\,{g}^{5}- 109478969.7\,{g}^{
13}$$ $$+ 953696332.6\,{g}^{15}- 4928124568.0\,{g}^{17}
$$ $$\beta_{{17}}=44638767880.0\,{g}^{18}+ 6126484.834\,{g}^{10}- 82996.74004\,{g}^{6}
$$ $$-1336.249746\,{g}^{4}+ 0.1005384968\,{g}^{2}+ 971831675.9\,{g}^{14}
$$ $$+27515.67004\,{g}^{8}- 88542747.03\,{g}^{12}- 8512571797.0\,{g}^{16}$$

\vfill\eject
The solutions for $Z$ and $g_0$ lead to the series expansion of the the logarithmic derivative of $Z$,
$\gamma=\sum_{n}\gamma_n/(\mu s)^n$, with the first 17 coefficients given by

$$\gamma_{{1}}= 0.3580986219\,g
$$ $$\gamma_{{2}}=- 0.3324601337\,{g}^{2}
$$ $$\gamma_{{3}}=- 0.1989436788\,g+ 0.7407774312\,{g}^{3}
$$ $$\gamma_{{4}}=- 2.157405193\,{g}^{4}+ 0.3591976683\,{g}^{2}
$$ $$\gamma_{{5}}=- 1.514499301\,{g}^{3}+ 7.609034060\,{g}^{5}+ 0.1566681471\,g
$$ $$\gamma_{{6}}=- 30.90102012\,{g}^{6}+ 6.324551298\,{g}^{4}- 0.2470671327\,{g
}^{2}
$$ $$\gamma_{{7}}= 2.810587640\,{g}^{3}+ 140.6420098\,{g}^{7}- 29.47246351\,{g}^
{5}$$ $$- 0.1342869832\,g
$$ $$\gamma_{{8}}=- 15.56377334\,{g}^{4}- 0.4393122678\,{g}^{2}- 705.4720543\,{g
}^{8}$$ $$+ 149.9640642\,{g}^{6}
$$ $$\gamma_{{9}}= 3856.625518\,{g}^{9}+ 0.1196770568\,g- 5.755944500\,{g}^{3}$$ $$-
 825.7869756\,{g}^{7}+ 92.59293353\,{g}^{5}
$$ $$\gamma_{{10}}= 3.423764699\,{g}^{2}+ 4887.996878\,{g}^{8}- 573.0603460\,{g}
^{6}$$ $$+ 38.04564185\,{g}^{4}- 22794.88025\,{g}^{10}
$$ $$\gamma_{{11}}=- 0.1091081738\,g- 30947.00771\,{g}^{9}+ 144798.2478\,{g}^{11
}$$ $$+ 15.15984462\,{g}^{3}+ 3727.952024\,{g}^{7}- 299.6137729\,{g}^{5}
$$ $$\gamma_{{12}}= 208764.5292\,{g}^{10}- 983835.0583\,{g}^{12}- 15.34219367\,{
g}^{2}$$ $$- 25536.87550\,{g}^{8}+ 2257.479359\,{g}^{6}- 75.61099893\,{g}^{
4}
$$ $$\gamma_{{13}}= 0.1009775167\,g+ 184253.1860\,{g}^{9}- 1495814.945\,{g}^{11}
$$ $$- 56.87562475\,{g}^{3}- 17294.46868\,{g}^{7}+ 1035.456051\,{g}^{5}$$ $$+
 7122090.122\,{g}^{13}
$$ $$\gamma_{{14}}= 61.55740610\,{g}^{2}+ 136372.4752\,{g}^{8}- 9808.072105\,{g}
^{6}$$ $$- 54746191.69\,{g}^{14}- 213.8081010\,{g}^{4}- 1399600.487\,{g}^{
10}$$ $$+ 11353026.88\,{g}^{12}
$$ $$\gamma_{{15}}=- 0.09445939125\,g- 3285.663980\,{g}^{5}- 91057448.51\,{g}^{
13}$$ $$+ 445520492.1\,{g}^{15}- 1114679.686\,{g}^{9}+ 11183177.87\,{g}^{11
}$$ $$+ 286.2007410\,{g}^{3}+ 90432.28371\,{g}^{7}
$$ $$\gamma_{{16}}= 42485.83958\,{g}^{6}+ 6557.967853\,{g}^{4}+ 9477782.019\,{g}
^{10}$$ $$- 93893690.17\,{g}^{12}- 3828085628.0\,{g}^{16}+ 770074284.4\,{g}
^{14}$$ $$- 238.8640816\,{g}^{2}- 829515.1962\,{g}^{8}
$$ $$\gamma_{{17}}=- 514466.9761\,{g}^{7}- 83973087.42\,{g}^{11}+ 0.08907659505\,g
$$ $$-1685.504320\,{g}^{3}+ 1614.901156\,{g}^{5}+ 827365423.6\,{g}^{13}
$$ $$-6852797795.0\,{g}^{15}+ 7721868.666\,{g}^{9}+ 34643996560.0\,{g}^{17}$$



\begin{thebibliography}{88}


\bibitem{Hat} B.Hatfield, {\it Quantum Field Theory of Particle and
Strings}, Addison Wesley, 1992, ISBN 0-201-11-11982X.

\bibitem{lusch} M.L\"uscher, R.Narayanan, P.Weisz and U.Wolff, Nucl. Phys. B384 (1992) 168.

\bibitem{Luscher} K.Jansen, C.Liu, M.Luscher, H.Simma,
S.Sint, R.Sommer, P.Weisz, U.Wolff, Phys. Lett. B372 (1996) 275.


\bibitem{Sint} S.Sint, Nucl. Phys. B451 (1995) 416.

\bibitem{Sint2} S.Sint, R.Sommer, Nucl. Phys.B465 (1996) 71.

\bibitem{Jac} R.Jackiw, {\it Analysis on Infinite Dimensional 
Manifolds:
Schrodinger
Representation for Quantized Fields}, Brazil Summer School 
1989:78-143.

\bibitem{Hugh} D.M.McAvity and H.Osborn, Nucl. Phys. B394 (1993) 
728.

\bibitem{REV2} J.H.Yee, {\it Schrodinger Picture Representation of
Quantum Field Theory}, Mt. Sorak Symposium 1991:210-271.

\bibitem{REV3} R.Floreanini, {\it Applications of Schrodinger Picture 
in
Quantum Field Theory}, Luc Vinet (Montreal U.), 
Edmonton Summer Inst.1987 v.2:53 (QC174.45:C2:1987) 


\bibitem{Kief1} C.Kiefer, Phys. Rev. D 45 (1992) 2044.

\bibitem{Kief2} C.Kiefer and A.Wipf, Ann. Phys. 236 (1994)241.


\bibitem{Feyn} R.P.Feynman, Nucl. Phys. B188(1981) 479.


\bibitem{MMaeda} M.Kawamura, K.Maeda, M.Sakamoto, Prog.Theor.Phys.97:939-948,1997


\bibitem{IS1} G.V.Dunne, R.Jackiw, C.A.Trugenberger , {\it 
Chern-Simons
Theory in the Schrodinger Representation}, Ann.Phys.194:197,1989.

\bibitem{IS2} A.V.Ramallo, {\it Two Dimensional Chiral Gauge 
Theories in
the Schrodinger
Representation}, Int.J.Mod.Phys.A5:153,1990.


\bibitem{else}  K.Heck, {\it  Some Considerations on the Problem of
Renormalization of Quantum
Field Theory in the Schrodinger Representation}. In German, Heidelberg
Univ. - HD-THEP 82-04 .


\bibitem{Wilson} K.G.Wilson, {\it Confinement of Quarks}
Phys.Rev.D10:2445-2459,1974 


\bibitem{Jeff} J.Greensite,  Nucl.Phys. B158 (1979) 469,
Nucl.Phys. B166 (1980) 113.



\bibitem{Halpern} M.B.Halpern, {\it Field Strength and Dual Variable
Formulations of Gauge Theory} Phys.Rev.D19 (1979) 517.

\bibitem{Iwa} J.Greensite, and J.Iwasaki, Phys. Lett. B223 (1989) 207.


\bibitem{Horiguchi2} T.Horiguchi, Nuovo. Cim. 111B (1996) 49,85,293.

\bibitem{HMS} T.Horiguchi, K. Maeda, M. Sakamoto, Phys.Lett. B344 
(1994) 105.

\bibitem{JK} J.Kowalski-Glikman, K.A.Meissner, Phys.Lett. B376 
(1996) 48.

\bibitem{JK2} J.Kowalski-Glikman, gr-qc/9511014.

\bibitem{JK3} A.Blaut, J.Kowalski-Glikman, gr-qc/9607004.

\bibitem{Maeda} K.Maeda, M.Sakamoto, Phys.Rev. D54 (1996) 1500.



\bibitem{Paul} P.Mansfield, Nucl. Phys. B418 (1994) 113.


\bibitem{Teper} B.Lucini, M.Teper, {\it SU(N) Gauge Theories
in (2+1)-Dimensions: Further Results} Phys. Rev. D 66 (2002) 09750.


\bibitem{Nair} D.Karabali, C-J.Kim, and V.P.Nair, {\it On the Vacuum
Wave Function and String Tension of Yang-Mills in (2+1)-Dimensions}
Phys. Lett. B 434:(1998) 103-109.



\bibitem{Sym} K.Symanzik, {\it Schrodinger Representation in
Renormalizable Quantum Field Theory},
Nucl. Phys. B190 [FS3] (1983) 1,
Les Houches 1982, Proceedings, Recent Advances In Field Theory and
Statistical Mechanics.




\bibitem{Cardy} J.Cardy, {\it Conformal Invariance} in {\it Phase Transitions and Critical Phenomena, Vol. 11} 55-126.
ed Domb, L. and Lebowitz, J



\bibitem{Paul1} P.Mansfield, Phys. Lett. B358 (1995) 287.

\bibitem{mo} T.R.Morris, J.F.Tighe, {\it Convergence of derivative
expansions in scalar field theory}, 
Int. J. Mod. Phys. A, Vol 16 (2001) 2095-2100.


\bibitem{Itzyk} I.Itzykson, J.-M.Drouffe, {\it Statistical Field Theory}
Cambridge University Press, 1989.

\bibitem{Par} G.Parisi, {\it Statistical Field Theory} Addison-Wesley, 1988.


\bibitem{Marcos} P.Mansfield, M.Sampaio, J.Pachos, 
{\it Short Distance Properties from Large Distance Behaviour} 
Int. J. Mod. Phys. A, Vol 13, 23 (1998) 4101-4122.


\bibitem{Shp} H.W.Diehl, M.Shpot, {\it Massive Field-Theory
approach to Surface Critical Behaviour in Three-Dimensional Systems},
Nucl.Phys. B528 (1998) 595-647.


\bibitem{Marcos2} P.Mansfield, M.Sampaio,{\it Yang-Mills 
Beta Function from a Large Distance Expansion of the \Sc Functional}
Nucl.Phys.B545 (1999) 623-655.

\end{thebibliography}
\end{document}